\documentclass[12pt]{article}

\usepackage{epsfig}
\usepackage{graphicx}
\usepackage{amsmath}
\usepackage{amssymb}
\usepackage{latexsym}
\usepackage{color}
\usepackage{mathrsfs}
\usepackage{natbib}
\usepackage{hyperref}
\usepackage{appendix}
\usepackage{dsfont}
\usepackage[left=3cm,top=4cm,right=3cm,bottom=3cm]{geometry}
\RequirePackage[caption=false]{subfig}
\usepackage{lscape}
\usepackage{dsfont}
\usepackage{rotating}

\newcommand{\de}{\mathrm{d}}

\newcommand{\cor}{\mathbb{C}\text{or}}

\newcommand{\s}{\mathbf{s}}
\newcommand{\timefreq}{u}
\DeclareMathOperator{\rr}{\triangledown}

\DeclareMathOperator{\T}{^{\mathsf T}}

\makeatletter
\newcommand*\bdot{\mathpalette\bdot@{.7}}
\newcommand*\bdot@[2]{\mathbin{\vcenter{\hbox{\scalebox{#2}{$\m@th#1\bullet$}}}}}
\makeatother
\def\V{\mathcal{V}}
\def\E{\mathcal{E}}
\def\G{\mathcal{G}}

\def\1{\mathbb{1}}

\newcommand{\R}{{\mathbb R}}

\newcommand{\indep}{\mathrel{{\perp}\hspace*{-0.6em}{\perp}}}

\newcommand{\given}{\mathrel{|}}
\newcommand{\condindep}[3]{#1 \indep #2 \given #3}

\DeclareSymbolFont{matha}{OML}{txmi}{m}{it}
\DeclareMathSymbol{\varv}{\mathord}{matha}{118}
\graphicspath{{./figs/}}

\begin{document}
\title{\LARGE \bf Graphical modelling and partial characteristics for multitype and multivariate-marked spatio-temporal point processes}
\date{\date{}}
\maketitle
\begin{center}
{{\bf Matthias Eckardt}$^{\text{a}}$, {\bf Jonatan A. Gonz{\'a}lez}$^{\text{b}}$ and {\bf Jorge Mateu}$^{\text{b}}$}\\
\noindent $^{\text{a}}$ School of Business and Economics, Humboldt Universit\"{a}t zu Berlin, Berlin, Germany\\
\noindent $^{\text{b}}$ Department of Mathematics, University Jaume I, Castell\'{o}n, Spain\\
\end{center}

\bigskip
\begin{abstract}

This paper contributes to the multivariate analysis of marked spatio-temporal point process data by introducing different partial point characteristics and extending the spatial dependence graph model formalism. Our approach  yields a unified framework for different types of spatio-temporal data including both, purely qualitatively (multivariate) cases and multivariate cases with additional quantitative marks. The proposed graphical model is defined through partial spectral density characteristics, it is highly computationally efficient and reflects the conditional similarity among sets of spatio-temporal sub-processes of either points or marked points with identical discrete marks. The paper considers three applications, two on crime data and a  third one on forestry.

\end{abstract}

\noindent%

\section{Introduction}

Spatio-temporal point patterns, where a finite set of pairs of $\{(\mathbf{s}_i, t_i)\}_{i=1}^n$ with $\mathbf{s}_i \in W\subseteq \mathbb{R}^2$ and $t_i\in T\subseteq \mathbb{R}^+$ are the point location and the time of occurrence of the $i$-th event, respectively, have become ubiquitous in various scientific areas arising in a number of scientific fields, such as infectious disease epidemiology \citep{GabrielDiggle2009}, the study of tornado events \citep{Gonzalez:JRSSC},  fire dynamics \citep{Moller2010FIRE} or  seismography \citep{doi:10.1080/01621459.1988.10478560, ChoiHall1999}. In turn, an ever-increasing demand for efficient statistical techniques, which not only account for the spatio-temporal specificity of the data but also facilitate an easy-to-read interpretation, is continuously emerging. Although  some progress has been made in the development of spatio-temporal characteristics and models, and different classical point process statistics such as the $J$-function \citep{doi:10.1111/sjos.12123}, Ripley's $K$ and pair correlation function \citep{DiggleChetwynd1995,GabrielDiggle2009,  mollerghorbani2012, gabriel2013raey}, or local indicators of spatio-temporal association functions \citep{doi:10.1002/env.2463} have been extended to the spatio-temporal case, marked spatio-temporal point patterns, where additional qualitative (yielding so-called multitype or multivariate point processes) or quantitative information is available for each pair $\{(\mathbf{s}_i, t_i)\}_{i=1}^n$, have not been covered much in the literature. Thus there is an increasing need of efficient statistical techniques allowing for the investigation and analysis of such type of spatio-temporal point processes. For a  general review on different spatio-temporal point process statistics and models commonly used at present we refer the interested reader to \cite{gonzalezetal2016}.

To approach this limitation, the aim of this paper is to contribute to the analysis of multivariate spatio-temporal point processes, where locations and times for a set of different types of points, such as a collection of distinct tree species, is under study. In particular, we consider multivariate marked spatio-temporal point processes where both qualitative and quantitative marks are available for each single pair $\{(\mathbf{s}_i, t_i)\}_{i=1}^n$.

At present, two different approaches can be identified in the literature focussing on quantitatively marked spatio-temporal point processes where one, potentially time-varying,  real-valued mark is attached to each single point location.
One strand of the literature, mainly applied in the field of spatio-temporal earthquake research, was covered by \cite{10025364994}, \cite{doi:10.1111/1467-9876.00420},  \cite{ChoiHall1999} and \cite{Marsan1076}) amongst others. They use \textit{mechanistic models} \citep{diggle2013book} which are defined through a parametric conditional intensity framework and express the infinitesimal expected rate of events at a particular time $t_i$ at point location $\mathbf{s}_i$ conditional on the history of the complete spatio-temporal  process up to time $t_i$ (see \cite{VereJones2009}). While the conditional intensity formalism provides a complete description of the process, the conditional intensity itself may be intractable or can not be evaluated exactly without numerical methods. Different from the mechanistic modelling approach, \cite{SARKKA20061698}, \cite{Renshaw2008}, \cite{Comas2009}, \cite{RENSHAW2009}, \cite{CRONIE20112209}, \cite{CronieEtAl2012}, \cite{Comas2013}, and \cite{REDENBACH2013672} discussed so-called \textit{growth-interaction processes} to model the evolution of a quantitatively marked spatial point pattern over equidistant steps in time such as the diameter at breast height (DBH) value for a set of trees recorded over consecutive times.

While the above specifications consider the analysis of quantitatively marked spatio-temporal point patterns, the investigation of cross-characteristics through  marked versions of the spatio-temporal reduced second-order moment measure and Ripley's $K$-function has just recently been discussed by \cite{Iftimi2019661}. Unlike the classical second-order summary characteristics such as the spatio-temporal $K$-function, the corresponding marked version allows to investigate the pairwise interrelation between subsets of points within one quantitatively marked spatio-temporal point pattern, e.g. the pairwise distance between juvenile and adult trees classified subject to a given threshold computed from the quantitative mark itself. That is, the marked spatio-temporal version of Ripley's $K$-function describes the expected number of further space-time points of type $j$ from an arbitrary space-time point of type $i$ of the process, given that the points in question have space and time separation $r \geq 0$ and $t\geq 0$, respectively.  Besides  these marked spatio-temporal point process characteristics, these authors also shortly pointed to extensions of classical cross- and dot-type point process characteristics for the multivariate case over equidistant steps in time in the supplementary material of their paper. Originating in the purely spatial case, these two types of point process characteristics for qualitatively marked processes investigate the pairwise distances between the point locations of two distinct component patterns or between the point locations of one component and those of any alternative patterns.

Different from the above approaches, the interest of the present paper are pairwise as well as global structural interrelations between different spatio-temporal  components conditional on all remaining components defined in terms of partial spatio-temporal point process characteristics and the spatio-temporal dependence graph model, respectively. While spatio-temporal point process characteristics quantify the conditional interrelation between two distinct components given all remaining components of the spatio-temporal process, the spatio-temporal dependence graph model simultaneously  elicits the potential structural interrelations between all spatio-temporal components in form of an undirected graph, and thus allows to detect directed and also induced spatio-temporal interdependencies. That is, our focus is the extension of the more classical concepts of partial correlation into the field of multivariate and multivariate-marked spatio-temporal point processes. Building upon the results of \cite{Eckardt2016},  \cite{Eckardt2019} and \cite{Eckardt2018partial}, this paper introduces different partial point process characteristics in the frequency domain. In addition, adopting the ideas of classical multitype point process characteristics, a new dot-type spectra, the {\em dot-spectra}, is introduced which reflects the linear interrelation of one component and any alternative patterns included.

To the best of our knowledge, the treatment of a combination of discrete with quantitative marks in a context of spatio-temporal point processes is new. If, in addition, we consider partial characteristics we go a step further with respect to the existing literature. Finally, the extension of a spatial dependence graph model to the spatio-temporal context is also new.

The remainder of the paper is structured as follows. Section 2 provides some background on the main characteristics of point processes in the spatio-temporal domain. Section 3 develops the main results of the spectral analysis for spatio-temporal point processes. Then, Section 4 presents the spatio-temporal dependence graph model. Applications to crime data and forestry are developed in Sections 5 and 6. The paper ends with some final conclusions.

\section{Spatio-temporal point processes}

To introduce spatio-temporal point processes, we follow \cite{gonzalezetal2016} and references therein as well as \cite{Iftimi2019661}. A spatio-temporal point process $X$ is, rigorously speaking, a random countable measure defined in a subset $W\times T$ of $\mathbb{R}^2 \times \mathbb{R}$ such that, for bounded $A\times B \subseteq W\times T$, $|X \cap(A\times B)|$ is finite, where $T$ is an interval in $\R_+ = [0,+\infty)$. A realisation of a point process is called a \textit{point pattern} and it is a finite set of pairs where the components are intended to state the spatial location $\mathbf{s}_i \in W$, and the time associated with that spatial location $t_i \in T$. Let $N(A\times B)$ denote the number of points of the set $(A\times B )\cap X$.

Stationarity and isotropy for spatio-temporal point processes can be defined as follows. $X$ is called {\em (spatio-temporally) stationary} when the process $(\mathbf{s},t) + X$ keeps the distribution of the original process $X$. On the other hand, $X$ is {\em (spatially) isotropic} if, for any rotation ${\bf r}$ around the origin, the rotated point process ${\bf r} X=\{({\bf r}\mathbf{s},t) : (\mathbf{s},t)\in X\}$ keeps the distribution of $X$.

Given a finite point process, it is frequently convenient to project it onto the spatial and temporal windows, so we can treat separately space and time \citep{mollerghorbani2012},
\begin{equation*}
X_{\text{space}} = \left\{\mathbf{s}:(\mathbf{s},t)\in X, t\in T\right\},
\quad
X_{\text{time}} = \left\{v:(\mathbf{s},t)\in X, \mathbf{s}\in W\right\}.
\end{equation*}

For the marked case consider the process $\{(\mathbf{s}_i, t_i), m_i\}_{i=1}^n$ where $m_i$ is a mark in a suitable mark space $\mathfrak{M}$. It is then called a spatio-temporal marked point pattern. When $\mathfrak{M} = \{1,2,\ldots, k\},k\ge 2,$ the process is called a \textit{multitype} spatio-temporal point process. The associated spatio-temporal point process is called the \textit{ground process}, and it is denoted by $X_{\text{g}}$.

\subsection{Spatio-temporal point process descriptors}
In analogy with the classical theory of random variables, we would like to deal with the distribution of the points of $X$ in $W\times T \times M$ where $M \subset \mathfrak{M}$. The {\em product densities} $\lambda^{(k)}$, $k\geq1$ describe the probability that there is a point of the process in each of the pairwise disjoints balls with centres in $k$ given points $\xi_1,\ldots,\xi_k$ and infinitesimal spatio-temporal marked volumes $\de \xi_1,\ldots,\de \xi_k\subseteq W\times T \times M$ with $\de \xi_i=\de \mathbf{s}_i\times \de t_i \times \nu(\de m_i)$, where $\nu()$ is a bounded reference measure on the mark space, and size $|\de \xi_i| = \de \mathbf{s}_i \de t_i \nu(\de m_i)$, $i=1,\ldots,k$. They can be defined by using Campbell's formula, which states that given a marked spatio-temporal point process $X$, for any non-negative function $h$ on $(\R^2\times\R \times \mathfrak{M})^k$,
\begin{equation}
\small{
	\label{ExpProdDens}
	\mathbb{E}\left[
	\mathop{\sum\nolimits\sp{\ne}}_{\xi_1,\ldots,\xi_k\in X}
	h(\xi_1,\ldots,\xi_k)
	\right]
	=
	\int \cdots \int h(\xi_1,\ldots,\xi_k)
	\lambda^{(k)}(\xi_1,\ldots,\xi_k) \de \xi_1\cdots \de \xi_k},
\end{equation}
where $\sum^{\ne}$ indicates that the summation is over distinct $k$-tuples of marked spatio-temporal events.

\subsubsection{Intensity function}\label{subsectionintensityfunctions}
As a first particular case of Eq. \eqref{ExpProdDens}, we focus on the intensity measure and intensity function. Usually, the analysis of a spatio-temporal point pattern starts with the estimation and modelling of the intensity function as it rules the univariate distribution of $X$ in $W\times T\times M$.
Considering the so-called {\em intensity measure} given by
$$\mu(A\times B\times C)=\mathbb{E}[N(A\times B\times C)], A\times B\times C \subseteq W\times T\times M,$$
when $\lambda=\lambda^{(1)}$ exists, we have that
	\begin{eqnarray*}
		\mu(A\times B\times C) = \int_{A}\int_{B} \int_{C} \lambda(\mathbf{s},t,m) \de \mathbf{s} \de t\nu(\de m),
	\end{eqnarray*}
and we call $\lambda(\mathbf{s},t,m)$ the {\em first-order intensity function} of $X$. Consider the projection of the process $X$ to only its spatio-temporal coordinates, the resulting process is called the \textit{ground process} and it is denoted by $X_{\text{g}}$. It can be shown that the intensity satisfies
$$\lambda(\mathbf{s},t,m)=f(m)\lambda_{\text{g}}(\mathbf{s},t),$$
where $\lambda_{\text{g}}(\mathbf{s},t)$ is the intensity of the ground process and $f(m)$ is a conditional density on $\mathfrak{M}$ in the spatio-temporal location $(\mathbf{s},t)$. In case that $X_{\text{g}}$ is stationary, or equivalently \textit{homogeneous}, then $\lambda_{\text{g}}(\mathbf{s},t)\equiv\lambda>0$. This constant is called the {\em intensity} of the ground process and $X$ is said \textit{spatio-temporally homogeneous. }

The first-order intensity of the ground process can be defined as well as
\begin{equation}\label{digglefirstorder}
\lambda_{\text{g}}(\mathbf{s},t)=\lim\limits_{|\de \mathbf{s}|,|\de t| \rightarrow 0}\frac{\mathbb{E} \left[ N(\de \mathbf{s}\times \de t) \right ]}{|\de \mathbf{s}| |\de t|}.
\end{equation}

When the first-order intensity function of the ground process $\lambda_{\text{g}}(\mathbf{s},t)$ can be factorised as
\begin{equation}\label{firstorderseparability}
\lambda_{\text{g}}(\mathbf{s},t) = \lambda_1(\mathbf{s}) \lambda_2(t),
\end{equation}
where
$\lambda_1(\cdot)$ and $\lambda_2(\cdot)$ are non-negative functions, then the process is called {\em first-order spatio-temporal separable}.  This separability is often taken as a working assumption in the literature in order to facilitate the estimations. In that case, the effects that are non-separable could be interpreted as second-order effects. Note that a stationary spatio-temporal point process $X$ is trivially first-order separable as its intensity is constant.

Once the sets $X_{\text{space}}$ and $X_{\text{time}}$ have been defined, it is naturally possible to define the marginal spatial and temporal intensity functions $\lambda_{\text{space}}$ and $\lambda_{\text{time}}$ as
\begin{equation*}
\lambda_{\text{space}}(\mathbf{s})
=
\lambda_{1}(\mathbf{s}) \int_{T}\lambda_2(t)\de t
\quad
\text{and}
\quad
\lambda_{\text{time}}(t)
=
\lambda_2(t) \int_{W}\lambda_1(\mathbf{s})\de \mathbf{s},
\end{equation*}
so that $\lambda_{\text{g}}(\mathbf{s},t) \propto \lambda_{\text{space}}(\mathbf{s}) \lambda_{\text{time}}(t)$, with $\lambda_{\text{g}}$, $\lambda_{\text{space}}$, $\lambda_{\text{time}}$ all being constant when $X$ is homogeneous.

To estimate the spatio-temporal first-order intensity function of the ground process $X_{\text{g}}$, the estimation of the marginal spatial and temporal intensity functions is first presented. For the spatial intensity function, a non-parametric specification which is most frequently used at present is defined in the form of a kernel estimator
\begin{equation*}
\hat{\lambda}_{\text{space}}\left(\mathbf{s}\right) =\sum_{i=1}^{n}\frac{k_{\epsilon}\left(\mathbf{s}-\mathbf{s}_{i}\right) }{c_{ \epsilon} \left(\mathbf{s}_{i};W\right) },\qquad \mathbf{s}\in W,
\label{estimationofdensityst1}
\end{equation*}
where
\begin{equation*}
k_{\epsilon }\left( \mathbf{s}\right) =\frac{1}{\epsilon^2}k\left( \frac{\mathbf{s}}{\epsilon }\right),
\end{equation*}
$k(\cdot)$ is a bivariate kernel and $\epsilon >0$ is the \textit{bandwidth}, and
\begin{equation*}
c_{\epsilon}\left(\mathbf{s}_{i}; W\right) =\int_{W}k_{\epsilon }\left(\mathbf{s}-\mathbf{s}_{i}\right) \de\mathbf{s}
\end{equation*}
is an edge-correction intended to stabilise the mass of the estimator so that its integral is roughly the number of points $n$. The marginal temporal intensity function $\lambda_{\text{time}}\left( t\right)$ can be estimated in the very same non-parametric fashion. The bandwidth is a sensitive parameter extremely delicate to be chosen; however, there are several methods to approach to a proper value. We note that there are some alternatives to estimate the spatial and temporal intensity components by using parametric or semi-parametric methods. The suitability of these approaches depends on how well we know the data if there are helpful covariates.

We note that under separability, given two unbiased estimators $\hat{\lambda}_{\text{space}}(\cdot)$ and $\hat{\lambda}_{\text{time}}(\cdot)$,  an unbiased estimator of the spatio-temporal first-order intensity function of the ground process $X_{\text{g}}$ is
\begin{equation*}
\hat{\lambda}_{\text{g}}\left(\mathbf{s},t\right) =\frac{1}{n}\left(\hat{\lambda}_{\text{space}}\left( \mathbf{s}\right) \hat{\lambda}_{\text{time}}\left( t\right)\right).
\label{estimationintensityst0}
\end{equation*}

When the spatio-temporal separability is not fulfilled there are some options to properly estimate the intensity, for instance, a non-separable estimator is given by \cite{Gonzalez:JRSSC}

\begin{equation*}
\hat{\lambda}_{\text{g}}^{\text{NS}}(\mathbf{s},t)=\sum_{i=1}^{n}\frac{k_{\epsilon}^{2}(\mathbf{s}-\mathbf{s}_i)k_{\delta}^{1}(t-t_i)}{c_{\epsilon}(\mathbf{s}_i;W)c_{\delta}(t_i; T)},
\label{lambdaNS}
\end{equation*}
where $k_{\delta}^{1}$ is a one-dimensional Gaussian kernel with bandwidth $\delta$ and $c_{\delta}(v;T)$ is the analogous to Diggle's edge-correction for the temporal component.

\subsubsection{Marked versions of the product density and $K$-function}

The so-called reduced second-order moment measure or product density function corresponds to a particular case $(\lambda^{(2)})$ of the family of product densities defined through Eq. \eqref{ExpProdDens}. This function depends on two marked spatio-temporal variables $(\mathbf{s}_1,t_1,m_1)$ and $(\mathbf{s}_2,t_2,m_2)$ and takes the form
$$
\lambda^{(2)}((\mathbf{s}_1,t_1,m_1),(\mathbf{s}_2,t_2,m_2))=f^{(2)}(m_1,m_2)\lambda^{(2)}_{\text{g}}((\mathbf{s}_1,t_1),(\mathbf{s}_2,t_2)),
$$
where $\lambda^{(2)}_{\text{g}}((\mathbf{s}_1,t_1),(\mathbf{s}_2,t_2))$ is the product density of $X_{\text{g}},$ and $f^{(2)}(m_1,m_2)$ is the density of the conditional probability of two points having marks $m_1$ and $m_2$ given their spatio-temporal locations $(\mathbf{s}_1,t_1)$ and $(\mathbf{s}_2,t_2)$.

Considering the ground process, analogously to Eq. \eqref{digglefirstorder}, the second-order product density function (or second-order spatio-temporal intensity function) is defined as \citep{diggle2013book}
\begin{equation}\label{productdensity}
\lambda^{(2)}_{\text{g}}((\mathbf{s}_1,t_1),(\mathbf{s}_2,t_2))=\lim\limits_{|\de \mathbf{s}_1|,|\de \mathbf{s}_2 |,|\de t_1|, |\de t_2|\rightarrow 0}\frac{\mathbb{E} \left[ N(\de \mathbf{s}_1\times \de t_1) N(\de \mathbf{s}_2\times \de t_2)\right ]}{|\de \mathbf{s}_1||\de \mathbf{s}_2 ||\de t_1||\de t_2|}.
\end{equation}

An important summary statistic for marked spatio-temporal point process is the \textit{pair correlation function}. This can be defined as the standardised version (and far more useful) of the product density function,
\begin{equation*}
g((\mathbf{s}_1,t_1,m_1),(\mathbf{s}_2,t_2,m_2))=\frac{f^{(2)}(m_1,m_2)}{f(m_1)f(m_2)}\times g_{\text{g}}((\mathbf{s}_1,t_1),(\mathbf{s}_2,t_2)), \quad (\mathbf{s}_1,t_1),(\mathbf{s}_2,t_2) \in W\times T,
\label{pcfng}
\end{equation*}
where $g_{\text{g}}$ is the pair correlation function of the ground process and it is given by
\begin{equation*}
g_{\text{g}}((\mathbf{s}_1,t_1),(\mathbf{s}_2,t_2))=\frac{\lambda^{(2)}_{\text{g}}((\mathbf{s}_1,t_1),(\mathbf{s}_2,t_2))}{\lambda_{\text{g}}(\mathbf{s}_1,t_1)\lambda_{\text{g}}(\mathbf{s}_2,t_2)}, \quad (\mathbf{s}_1,t_1),(\mathbf{s}_2,t_2) \in W\times T.
\label{pcf}
\end{equation*}

The advantage of having a standardised product density function is that this function takes the constant value $1$ for a spatio-temporal complete random process in the presence of independent marking. So values above or below this benchmark will be easily interpreted towards clustering or regularity.

One of the most important working assumptions when dealing with marked spatio-temporal point processes is the concept of \textit{second-order intensity-reweighted stationarity} defined as follows. A marked spatio-temporal point process $X$ is second-order intensity-reweighted stationary \textit{(SOIRS)} \citep{GabrielDiggle2009} if
\begin{equation*}
	g((\mathbf{s}_1,t_1,m_1),(\mathbf{s}_2,t_2,m_2))=\bar{g}(\mathbf{s}_1-\mathbf{s}_2,t_1-t_2,m_1,m_2),
\end{equation*}
for any $(\mathbf{s}_1,t_1),(\mathbf{s}_2,t_2) \in W\times T$, where $\bar{g}$ is a non-negative function.
	
If the process is also isotropic, $\bar{g}(\mathbf{s}_1-\mathbf{s}_2,t_1-t_2,m_1,m_2) = g_0 (r,t,m_1,m_2)$, meaning that the pair correlation depends only on the distances $r = \|\mathbf{s}_1-\mathbf{s}_2\|$ and $t = |t_1-t_2|$, where $g_0$ is a non-negative function.

One of the most common methods for the estimation of the pair correlation function is the non-parametric kernel approach since such estimator is easy to interpret and implement. Assuming that the spatio-temporal point pattern is given by a sequence of pairs $X_{\text{g}}=\{(\mathbf{s}_{i},t_i)\}_{i=1}^n$, the estimator is given by
\begin{equation*}
\hat{g}_{\text{g}}(r,t)=\frac{1}{4\pi r}
\sum_{i=1}^n
\sum_{j\neq i}
\frac{k_{1 \epsilon}(\|\mathbf{s}_{i}-\mathbf{s}_{j}\|- r)k_{2 \delta}(|t_{i}-t_{j}|-t)}{\hat{\lambda}_{\text{g}} \left( \mathbf{s}_{i},t_{i}\right) \hat{\lambda}_{\text{g}} \left(\mathbf{s}_{j},t_{j}\right) w_{ij}
}, \quad r> \epsilon,t>\delta,
\label{gstestimada1}
\end{equation*}
where $k_{1 \epsilon}$ and $k_{2 \delta}$ are kernel functions with spatial and temporal bandwidths $\epsilon$ and $\delta$, and $w_{ij}$ is an edge-correction factor for correcting the lack of information occurring between points close to the edge of $W\times T$ and the unobserved outsider points (see e.g,\ \citealp{gabriel2013raey}).

A spatio-temporal adaptation of the mean product of marks sited a distance $r$ apart (see e.g, \citealp{renshaw2002spectra}) can be thought of as a natural extension by including the temporal dimension, i.e, for a stationary and isotropic process
$$
U(r,t) = \lambda^2 g_{\text{g}}(r,t)S(r,t)\de\mathbf{s}_1 \de t_1 \de\mathbf{s}_2 \de t_2,
$$
where $\de\mathbf{s}_1$ and $\de\mathbf{s}_2$ are two infinitesimal spatial areas separated by a distance $r$ and $\de t_1$ and $\de t_2$ are two infinitesimal temporal lengths separated by a distance $t$. $S(r,t)$ represents a spatio-temporal mark correlation function that has not yet been examined in the current literature and that deserves especial attention given its extremely usefulness for analysing complex point patterns.

Finally, the marked spatio-temporal $K$-function was defined in \cite{Iftimi2019661} in its general version. We can take advantage of the pair correlation function in the case of SOIRS processes. Let $C,D\subset \mathfrak{M}$, so the $K$-function is given by
\begin{equation*}
K^{CD}(r,t)=\frac{1}{\nu(C)\nu(D)}\int_{C}\int_{D} \int_{||\mathbf{s}||\leq r}\int_{-t}^{t}g((\mathbf{0}, 0, m_1),(\mathbf{s},v,m_2))\de \mathbf{s} \de v \nu(\de m_2) \nu(\de m_1).
\end{equation*}
For Poisson processes, the $K$-function is $2\pi r^2t$. This statistic can be estimated through the following expression
\begin{equation*}
\hat{K}^{CD}(r,t)=
\sum_{i=1}^n
\sum_{j\neq i}
\frac{\mathbf{1}(\|\mathbf{s}_{i}-\mathbf{s}_{j}\|\leq r)\mathbf{1}(|t_{i}-t_{j}|\leq t)\mathbf{1}(m_i\in C)\mathbf{1}(m_j\in D)}{\hat{\lambda} \left( \mathbf{s}_{i},t_{i},m_i\right) \hat{\lambda} \left(\mathbf{s}_{j},t_{j},m_j\right) \nu(C)\nu(D) w_{ij}},
\label{kstestimada1}
\end{equation*}
where $\mathbf{1}(\cdot)$ is the indicator function, and $w_{ij}$ is a suitable edge-correction.

\subsubsection{Multitype spatio-temporal point patterns}

In analogy with the classical theory of multitype point patterns, we can define some useful descriptors. Consider a multitype spatio-temporal point process composed by $q$ types of points, so that $X=\bigcup_{i=1}^q X^{(i)}$. It is straightforward to consider a multitype point process as a marked point process where $\mathfrak{M}$ is the set of indices $\{1,\ldots,q\};$ thus, the measure $\nu(\cdot)$ is the counting measure.

Let $N_i(A\times B)$ denote the number of points of type $i$ of the set $(A\times B )\cap X$. We can naturally define the \textit{spatio-temporal cross-product density function} as
\begin{equation}
\lambda^{(2)}_{ij}((\mathbf{s}_1,t_1),(\mathbf{s}_2,t_2))=\lim\limits_{|\de \mathbf{s}_1|,|\de \mathbf{s}_2 |,|\de t_1|, |\de t_2|\rightarrow 0}\frac{\mathbb{E} \left[ N_i(\de \mathbf{s}_1\times \de t_1) N_j(\de \mathbf{s}_2\times \de t_2)\right ]}{|\de \mathbf{s}_1||\de \mathbf{s}_2 ||\de t_1||\de t_2|}.
\end{equation}
Note that $\lambda^{(2)}_{ii} \equiv \lambda^{(2)}$ in Eq. \eqref{productdensity}. The \textit{spatio-temporal cross-covariance density function} for, say, type $i$-points can be defined by
\begin{equation}\label{eq:crosscov}
\zeta_{ij}((\mathbf{s}_1,t_1),(\mathbf{s}_2,t_2)) =\lambda^{(2)}_{ij}((\mathbf{s}_1,t_1),(\mathbf{s}_2,t_2))-\lambda_i(\mathbf{s}_1,t_1)\lambda_j(\mathbf{s}_2,t_2).
\end{equation}
Having the first- and second-order characteristics at hand, two alternative statistics can be defined: (a) the spatio-temporal correlation function
\[
\cor((\mathbf{s},t),(\mathbf{s}^\prime,t^\prime))=(d\mathbf{s}dtd\mathbf{s}^\prime dt^\prime)^{1/2}\frac{\zeta_{ij}((\mathbf{s},t),(\mathbf{s}^\prime,t^\prime))}{(\zeta_{i}((\mathbf{s},t),(\mathbf{s}^\prime,t^\prime))\zeta_{j}((\mathbf{s},t),(\mathbf{s}^\prime,t^\prime)))^{1/2}}
\]
and (b) the scaled cross-covariance density function
\[
\tau_{ij}((\mathbf{s},t),(\mathbf{s}^\prime,t^\prime))=\frac{\zeta_{ij}((\mathbf{s},t),(\mathbf{s}^\prime,t^\prime))}{(\lambda_{i}(\mathbf{s},t)\lambda_{j}(\mathbf{s}^\prime,t^\prime))^{1/2}}
\]

All these functions are well defined under some regularity conditions, indeed multiple coincident events are precluded. This assumption implies that
$$\mathbb{E} \left[ N^2(\de \mathbf{s}\times \de t)\right] = \mathbb{E} \left[ N(\de \mathbf{s}\times \de t)\right] = \lambda(\mathbf{s},t)|\de \mathbf{s}||\de t|,$$
which leads to a natural definition of \textit{Bartlett's complete covariance density function} as
\begin{equation}\label{stpp:completecov}
\small
\begin{aligned}
\kappa_{ii}(\mathbf{s}_1,\mathbf{s}_2,t_1,t_2)&=\lambda_i(\mathbf{s}_1,t_1)\delta(\mathbf{s}_1-\mathbf{s},t_1-t_2)+\zeta_{ii}(\mathbf{s}_1,\mathbf{s}_2,t_1,t_2)\\
 &= \lambda_i(\mathbf{s}_1,t_1)\delta(\mathbf{s}_1-\mathbf{s},t_1-t_2)+\lambda_{ii}^{(2)}((\mathbf{s}_1,t_1),(\mathbf{s}_2,t_2))-\lambda_i(\mathbf{s}_1,t_1)\lambda_i(\mathbf{s}_2,t_2)
\end{aligned}
\end{equation}
where $\delta(\cdot)$ is a  multivariate Dirac delta function with
\[
\delta(\mathbf{s}_1-\mathbf{s}_2,t_1-t_2)=
\begin{cases}
1 \text{~if~} \mathbf{s}_1-\mathbf{s}_2=\mathbf{0} \text{~and~} t_1-t_2=0\\
 0 \text{~otherwise}.
\end{cases}
\]
We note that this function simplifies under second-order stationarity to
$\kappa_{ii}(\mathbf{c},h)=\lambda_{ii}^{(2)}(\mathbf{c},h)-\lambda^2+\lambda_i\delta(\mathbf{c},h)$.

Generalising Bartlett's complete covariance density function to the spatio-temporal case, we assume that the spatio-temporal cross-covariance and the complete cross-covariance density functions coincide, such that
\begin{eqnarray*}
\kappa_{ij}(\mathbf{s}_1,\mathbf{s}_2,t_1,t_2)&=&\zeta_{ij}(\mathbf{s}_1,\mathbf{s}_2,t_1,t_2) \quad \text{and}\\
\kappa_{ji}(\mathbf{s}_1,\mathbf{s}_2,t_1,t_2)&=&\zeta_{ji}(\mathbf{s}_1,\mathbf{s}_2,t_1,t_2) \quad \text{for } i\neq j.
\end{eqnarray*}

\section{Spectral analysis of spatio-temporal point processes}
Extending \cite{Dorai-Raj2001} to the present context, and requiring orderliness and discrete equidistant points in time, we define the auto- and cross-spectral density functions of the $i$-th and $j$-th components of a multivariate second-order stationary spatio-temporal point process as the Fourier transform of the complete spatio-temporal auto- and cross-covariance density functions at frequencies $\mathbf{w}_{st}=(\mathbf{w}_\s,w_\timefreq$). By this, the auto-spectrum of the $i$-th component is
\begin{equation}\label{eq:autospec:stpp}
\begin{split}
f_{ii}(\mathbf{w}_\s,w_\timefreq)&=\int_{\mathbb{R}^2}\int_{\mathbb{R}}\kappa_{ii}(\mathbf{c},h)\exp\left(-\imath(\mathbf{w}_\s\T\mathbf{c}+w_\timefreq h)\right)dh d\mathbf{c}\\
&=\lambda+\int_{\mathbb{R}^2}\int_{\mathbb{R}}\zeta_{ii}(\mathbf{c},h)\exp\left(-\imath(\mathbf{w}_\s\T\mathbf{c}+w_\timefreq h)\right) dh d\mathbf{c},
\end{split}
\end{equation}
where $f_{ii}(\mathbf{w}_\s,w_\timefreq)=f_{ii}^{(\mathbf{s})}(\mathbf{w}_\s)\circ f_{ii}^{(t)}(w_\timefreq)$ is the convolution of the auto-spectral density functions for the spatial and temporal components, $\mathbf{w}_\s=( w_p, w_q)$ is a two-dimensional array of spatial frequencies and $w_\timefreq$ is a vector of temporal frequencies. Applying Bochner's theorem, $\kappa_{ii}(\mathbf{c},k)$ can be recovered by the inverse Fourier transform of  \eqref{eq:autospec:stpp},
\[
\kappa_{ii}(\mathbf{c},k)=\int_{\mathbb{R}^2}\int_{\mathbb{R}}\exp(\imath(\mathbf{w}_\s\T\mathbf{c}+w_\timefreq h))f_{ii}(\mathbf{w}_\s,w_\timefreq)dw_\timefreq d\mathbf{w}_\s.
\]
Further, under second-order spatio-temporal separability, \eqref{eq:autospec:stpp} simplifies and allows for the decomposition into
\[
f_{ii}(\mathbf{w}_\s,w_\timefreq)=\int_{\mathbb{R}^2}\exp(-\imath\mathbf{w}_\s\T\mathbf{c})\kappa_{ii}(\mathbf{c})d\mathbf{c}\int_{\mathbb{R}} \exp(-\imath w_\timefreq h)\kappa_{ii}(h)dh
\]
 which yields, by inverse Fourier operation, again the complete spatio-temporal auto-covariance density function in the form of
\[
\kappa_{ii}(\mathbf{c},k)=\int_{\mathbb{R}^2}\exp(\imath\mathbf{w}_\s\T\mathbf{c})f_{ii}^{(\mathbf{s})}(\mathbf{w}_\s)d\mathbf{w}_\s\int_{\mathbb{R}} \exp(\imath w_\timefreq h)f_{ii}^{(t)}(w_\timefreq)dw_\timefreq.
\]

Likewise, the spatio-temporal cross-spectral density function $f_{ij}(\mathbf{w}_\s,w_\timefreq)=f_{ij}^{(\mathbf{s})}(\mathbf{w}_\s)\circ f_{ij}^{(t)}(w_\timefreq),$ which measures the linear interrelation of the spatio-temporal components $N_i$ and $N_j$, is defined by
\begin{equation}\label{eq:crossspectra}
f_{ij}(\mathbf{w}_\s,w_\timefreq)=\int_{\mathbb{R}^2}\int_{\mathbb{R}}\exp(-\imath(\mathbf{w}_\s\T\mathbf{c}+w_\timefreq h))\kappa_{ij}(\mathbf{c},h) dh d\mathbf{c}.
\end{equation}
As $\zeta_{ij}(\mathbf{c},h)=\zeta_{ji}(-\mathbf{c},-h)$ and $\kappa_{ij}(\mathbf{c},h)=\kappa_{ji}(-\mathbf{c},-h)$ under second-order stationarity, we have $f_{ij}(\mathbf{w}_s,w_\timefreq)=f_{ji}(-\mathbf{w}_s,-w_\timefreq)$ such that it suffices to consider only one cross-spectrum. However, at the same time, as the spatio-temporal cross-covariance density function is not necessarily symmetric, i.e. $\zeta_{ij}(\mathbf{c},k)\neq\zeta_{ij}(\mathbf{-c},-k)$, $f_{ij}(\mathbf{w}_\s,w_\timefreq)$ is a complex-valued function which can be decomposed into its real and imaginary parts either in terms of Cartesian or polar coordinates yielding the co-spectrum $C_{ij}(\mathbf{w}_\s)$, the quadrature spectrum $Q_{ij}(\mathbf{w}_\s,w_\timefreq)$, the amplitude spectrum $\mathfrak{a}_{ij}(\mathbf{w}_\s,w_\timefreq)=\mod(f_{ij}(\mathbf{w}_\s,w_\timefreq))$ and the phase spectrum $\wp_{ij}(\mathbf{w}_\s,w_\timefreq)=\tan^{-1}\left(-Q_{ij}(\mathbf{w}_\s,w_\timefreq)/C_{ij}(\mathbf{w}_\s,w_\timefreq)\right)$. The amplitude spectrum represents the relative magnitude of the power attributable to  frequencies $(\mathbf{w}_\s,w_\timefreq)$ while the phase spectrum indicates the similarity of two patterns up to linear shifts (cf. \citet{ Chatfield1989, Priestley1981}).

Although the spatio-temporal cross-spectrum provides insights into the linear interrelation of two components at frequencies  $\mathbf{w}_{st}$, it is preferable to compute the spatio-temporal spectral coherence function,
\begin{equation}\label{eq:STPPCoh}
\vert R_{ij}(\mathbf{w}_{st})\vert^2=\frac{f_{ij}(\mathbf{w}_{st})^2}{\left[f_{ii}(\mathbf{w}_{st})f_{jj}(\mathbf{w}_{st})\right]},
\end{equation}
satisfying $0\leq \vert R_{ij}(\mathbf{w}_{st})\vert^2\leq 1$. For a subset of $J$ components, a different spectral coherence function which quantifies the extend to which the $i$-th component is determinable from the $J$ components is the {\em multiple coherence function} $\vert R^m_{iJ}(\mathbf{w}_{st})\vert^2$,
\begin{equation}\label{eq:MultCoh}
\vert R^m_{iJ}(\mathbf{w}_{st})\vert^2=\frac{f_{iJ}(\mathbf{w}_{st})f_{JJ}^{-1}(\mathbf{w}_{st})f_{Ji}(\mathbf{w}_{st})}{f_{ii}(\mathbf{w}_{st})}.
\end{equation}

Defining $\kappa_{i\bullet}(\mathbf{c},h)$ as the complete dot-type cross-covariance function between the $i$-th and all $\mathbf{N}\setminus \lbrace i \rbrace$-th components and substituting $\kappa_{i\bullet}(\mathbf{c},h)$ for $\kappa_{ij}(\mathbf{c},h)$ in \eqref{eq:crossspectra} yields the dot-type spatio-temporal cross-spectrum
\begin{equation}\label{eq:dotcrossspectra}
f_{i\bullet}(\mathbf{w}_\s,w_\timefreq)=\int_{\mathbb{R}^2}\int_{\mathbb{R}}\exp(-\imath(\mathbf{w}_\s\T\mathbf{c}+w_\timefreq h))\kappa_{i\bullet}(\mathbf{c},h)dh d\mathbf{c}
\end{equation}
from which, in turn, the dot-type spatio-temporal spectral coherence function $\vert R_{i\bullet}(\mathbf{w}_\s,w_\timefreq)\vert^2$ and multiple coherence function $\vert R^m_{iJ}(\mathbf{w}_{st})\vert^2$ can  be computed. We note that $\vert R^m_{iJ}(\mathbf{w}_{st})\vert^2$ and $\vert R_{i\bullet}(\mathbf{w}_{st})\vert^2$ coincide whenever $J$ equals $\mathbf{N}\setminus \lbrace i \rbrace$. Unlike the ordinary cross-spectral characteristics, the above dot-type functions express the linear interrelation between one particular component and the set of all remaining components.

Besides the spectral coherence functions, a different spectral quantity which measures the linear effect of the $j$-th on $i$-th (resp. $i$-th on $j$-th) component is the spatio-temporal gain spectrum $G_{i|j}(\mathbf{w}_\s,w_\timefreq)$ (resp. $G_{j|i}(\mathbf{w}_\s,w_\timefreq)$) defined by
\[
G_{i|j}(\mathbf{w}_\s,w_\timefreq)=\frac{\sqrt{f_{ii}(\mathbf{w}_\s,w_\timefreq)R_{ij}(\mathbf{w}_\s,w_\timefreq)}}{f_{jj}(\mathbf{w}_\s,w_\timefreq)}
\]
where $G_{j|i}(\mathbf{w}_\s,w_\timefreq)$ can be computed analogous to  $G_{i|j}(\mathbf{w}_\s,w_\timefreq)$. Defining a dot-type version of the above functions yields the expression
\[
G_{i|\bullet}(\mathbf{w}_\s,w_\timefreq)=\frac{\sqrt{f_{ii}(\mathbf{w}_\s,w_\timefreq)R_{i\bullet}(\mathbf{w}_\s,w_\timefreq)}}{f_{\bullet\bullet}(\mathbf{w}_\s,w_\timefreq)}
\]
where $f_{\bullet\bullet}$ is the auto-spectrum defined over all components except $i$. This function provides information on the linear effect of all the alternative components on the $i$-th component.

Analogous to  \eqref{eq:autospec:stpp}, the marked spatio-temporal auto-spectrum for the $i$-th component of a multivariate-marked point process is defined as the Fourier transform of the auto-type spatio-temporal mean product of marks $U_{ii}$, and is given by
\begin{equation}
\label{eq:fourierautoMMSPP}
f^m_{ii}(\mathbf{w}_\s,w_\timefreq) = \int_{\mathbb{R}^2}\int_{\mathbb{R}} U_{ii}(\cdot)\exp(-\imath(\mathbf{w}_\s\T\mathbf{c}+w_\timefreq h)dh d\mathbf{c}.
\end{equation}
The cross-term expression of \eqref{eq:fourierautoMMSPP} is  obtained in the same way through the  Fourier transform of $U_{ij}$. Recapitulating Bochner's theorem, we note that again both  quantities $U_{ii}$ and $U_{ij}$ can uniquely be recovered by the inverse Fourier operations. As for the multitype case, we note that the marked cross-spectrum could be extended to a dot-type version by substituting $U_{ij}$ by the dot-type spatio-temporal mean product of marks $U_{i\bullet}$. This statistic would then include dot-type versions of the spatio-temporal pair and mark correlation functions which need further rigorous investigations in future research.

Adopting the ideas of \cite{Renshaw1983,Renshaw1984}, two different spectral representations of the spatio-temporal (marked) spectra can be defined. These spectra, the $\mathcal{R}$- and the $\Theta$-spectrum at the temporal frequency $w_\timefreq$ can be computed directly by converting the spatial frequency $\mathbf{w}_\s$ into polar form $\boldsymbol{w}_{\rr\theta}$  with $\rr=\sqrt{p^2+q^2}$ and $\theta=\tan^{-1}(p/q)$. This yields the $\mathcal{R}_t$-spectrum $\hat{f}_{\mathcal{R}}(\rr,w_\timefreq)=\frac{1}{n_{\rr}}\sum_{\rr^\prime}\sum_\theta\hat{f}^{\circ}(\mathbf{w}_{\rr^\prime\theta},w_\timefreq),~\rr= 1,2,\ldots$ where $n_{\rr}$ represents the number of periodogram ordinates $(p,q)$ for which $1-\rr <\rr^\prime\leq \rr$ and the $\Theta$-spectrum $\hat{f}_\Theta(\theta,w_\timefreq)=\frac{1}{n_\theta}\sum_{\rr}\sum_{\theta^\prime}\hat{f}^{\circ}(\mathbf{w}_{\rr\theta^\prime},w_\timefreq),~\theta=0^{\circ}, 10^{\circ},\ldots, 170^{\circ}$ where   $n_\theta$ is the number of periodogram ordinates for which $\theta-5^{\circ}<\theta^\prime\leq\theta+5^{\circ}$, respectively. Here $\hat{f}^{\circ}$ denotes the polar form computed from the ordinary (marked) spatio-temporal cross-spectra at temporal frequency $w_\timefreq$. While the $\mathcal{R}$-spectrum provides useful information on the scales of point patterns under the assumption  of isotropy, the $\Theta$-spectrum could be used to investigate directional features of the point pattern.

\subsection{Estimation of spatio-temporal spectral density functions}

Next, the estimation of both functions from empirical data by means of spatio-temporal auto- and cross-periodograms is presented. Assume that we have a multivariate spatio-temporal point pattern in a bounded region $W\times T\subset \mathbb{R}^2\times\mathbb{R}^+$ where $W$ is required to be  a rectangular region with sides of length $l_1$ and $l_2$ and  independent of $t$.

Let $\{s_i(t)\} = \{(x_{ik}(t); y_{ik}(t))\}$ with $i=1,\ldots, n_i$ (resp $\{s_j(t)\}$)  denote the locations of points of type $i$ (resp. of type $j$) recorded at time $t,~t\in T$.  To ease notation, we will be using only the subindex $k$ for the coordinates, whenever no confusion arises. For simplicity, we assume that the locations have been scaled to the unit square prior to analysis and that $T$ is an ordered set of consecutive times recorded for equidistant steps in discrete time. 
The spatio-temporal auto- and cross-periodograms result from the DFT of the point locations $\lbrace \mathbf{s}_i(t)\rbrace$ and $\lbrace \mathbf{s}_j(t)\rbrace$ at time $t\in T$. For events of type $i$, the DFT is defined as
\begin{equation}\label{DFT:STPP1}
\begin{split}
\mathcal{F}_i(p,q,\timefreq)&=\sum^T_{t=1}\sum^{n_{i}}_{k=1}\exp\left(-2\pi\imath\left((px_k(t)+qy_k(t))+\timefreq t/T\right)\right)\\
&= a_i(p,q,\timefreq)+\imath b_i(p,q,\timefreq)
\end{split}
\end{equation}
where $a_i(p,q,\timefreq)$ and $b_i(p,q,\timefreq)$ are the real and the imaginary parts of $\mathcal{F}_i(p,q,\timefreq)$, $p=0,1,2,\ldots,~q=\pm 1, \pm 2, \ldots$  and $\timefreq=-\left[\frac{T-1}{2}\right],\ldots, \left[\frac{T}{2}\right]$. In general, $p$ and $q$ are assumed to be independent of $\timefreq$.

Under second-order separability, \eqref{DFT:STPP1} factorises to
\begin{equation}\label{DFT:STPPseparable}
\begin{split}
\mathcal{F}_i(p,q,\timefreq)&=\sum^T_{t=1}\exp\left(-2\pi\imath\left(\frac{\timefreq t}{T}\right)\right) \sum^{n_{i}}_{k=1}\exp(-2\pi\imath(px_k(t)+qy_k(t)))\\
&=\sum^T_{t=1}\exp\left(-2\pi\imath\left(\frac{\timefreq t}{T}\right)\right)\mathcal{F}^{(t)}_{i}(p,q)
\end{split}
\end{equation}
where $\mathcal{F}^{(t)}_{i}(p,q)$ is the Fourier transform of the spatial frequencies $( w_p, w_q)$  for events of type $i$ at time $t$. From this expression, the spatio-temporal auto-periodogram for frequencies $\mathbf{w}_\s=(2\pi p/n_i, 2\pi q/n_i)$ and $w_\timefreq=2\pi \timefreq$ itself follows as
\begin{equation}\label{formelstppautospec}
\begin{split}
\widehat{f}_{ii}(\mathbf{w}_\s,w_\timefreq)&=\mathcal{F}_i(p,q,\timefreq)\overline{\mathcal{F}}_i(p,q,\timefreq)\\
&=\left[\sum^T_{t=1}\exp\left(-2\pi\imath \left(\frac{\timefreq t}{T}\right)\right)F^{(t)}_{i}(p,q)\right]\times\left[\sum^T_{t^\prime=1}\exp\left(2\pi\imath  \left(\frac{\timefreq t^\prime}{T}\right)\right)\overline{F}^{(t^\prime)}_{i}(p,q)\right]\\
&=\sum^T_{t=1}\sum^T_{t^\prime=1}\mathcal{F}^{(t)}_{i}(p,q)\overline{\mathcal{F}}^{(t^\prime)}_{i}(p,q)\exp(\imath w_\timefreq h/T)
\end{split}
\end{equation}
where $h=t-t^\prime$ is the time lag and $\overline{\mathcal{F}}^{(t)}_i(\cdot)$ is the complex conjugate of $\mathcal{F}^{(t)}_i(\cdot)$.

The computation of the spatio-temporal cross-periodogram follows analogously to \eqref{formelstppautospec} leading to
\[
\widehat{f}_{ij}(\mathbf{w}_\s,w_\timefreq)=\mathcal{F}_i(p,q,\timefreq)\overline{\mathcal{F}}_j(p,q,\timefreq)
\]
where $p,q$ and $\timefreq$ are defined as above.

Likewise, in the presence of both one qualitative and one quantitative marks  for each point location, both the marked spatio-temporal auto- and the marked spatio-temporal cross-periodograms result from the discrete Fourier transforms of the marked locations $\lbrace\mathbf{s}_i(t), m_i(\mathbf{s}_i(t))\rbrace$ and $\lbrace\mathbf{s}_j(t), m_j(\mathbf{s}_j(t))\rbrace$  at time $t\in T$. Assuming that the marked locations have been scaled to the unit square prior to the analysis, the spatial component $\mathcal{F}^{(t)}_{i}(p,q)$ changes to
\begin{equation}\label{ed:DFTmark}
\mathcal{F}^m_i(p,q)=\left(\sum_{k=1}^{n_i}\left(m_{k}(x_k(t),y_k(t))-\mu(m_{k}(x_k(t),y_k(t))\right)\exp(-2\pi\imath(px_k+qy_k))\right)
\end{equation}
where $\mu(m_{k}(x_k(t),y_k(t))$ is the mean computed over all quantitative marks for the $i$-th component. Plugging-in this expression for $\mathcal{F}^{(t)}_{i}(p,q)$ into \eqref{formelstppautospec} yields 
\begin{equation}\label{ed:DFTmarkSTPP}
\begin{aligned}
\mathcal{F}^m_i(p,q,\timefreq)=&\sum^T_{t=1}\exp\left(-2\pi\imath\left(\frac{\timefreq t}{T}\right)\right)\sum_{k=1}^{n_i}\left(m_{k}(x_k(t),y_k(t))-\mu(m_{k}(x_k(t),y_k(t))\right)\times\\
&\exp(-2\pi\imath(px_k(t)+qy_k(t)))
\end{aligned}
\end{equation}
where $p, q$  and $\timefreq$ are defined as above.

\section{Spatio-temporal dependence graph model}\label{Sec:stdg}
Although some progress has been made in the analysis of marked spatio-temporal point patterns and various point process characteristics can be found in the literature, the need for efficient exploratory techniques for multivariate and multivariate-marked spatio-temporal point patterns which allow for the simultaneous investigation of potential conditional interrelations among all component patterns still remains. To overcome this limitation, this section extends the framework of the spatial dependence graph model to introduce a new class of spatio-temporal dependence graph models which allows for the joint analysis of potential direct and indirect interrelations in multivariate and multivariate-marked spatio-temporal point patterns. In addition, using the same ideas underpinning the graphical model, different  partial spatio-temporal point process characteristics are introduced which represent the pair interrelation between to (marked) components conditional on all remaining patterns. To put it differently, we are interested in the partial linear interrelations between two component processes which remain conditional on all alternative components as expressed by  partial spatio-temporal  spectral characteristics, i.e. the partial spatio-temporal spectral density function $f_{ij|\V\backslash\lbrace i,j\rbrace}(\mathbf{w}_{st})$, which are next presented.

\subsection{Partial spatio-temporal spectral properties}\label{sec:partialSPP}
To formalise the concept of partial spectral properties, let $\mathbf{N}_\V$ denote a $d$-variate spatio-temporal point patterns indexed by $\V=1,\ldots,d$ where $d\geq 3$ and $N_{\V\backslash\lbrace i,j\rbrace}$ denote all alternative components of $\mathbf{N}_\V$ except $N_i$ and $N_j$.  Different from the ordinary spectral properties which do not help to distinguish between direct and induced interrelations, the objective of interest of this  section are linear interrelations between any pair of distinct components $(N_i,N_j)$ conditional on $N_{\V\backslash\lbrace i,j\rbrace}$. That is, the pairwise linear interrelation of $N_i$ and $N_j$ which remains after the linear effect of all alternative components has been removed. In this respect, the partial cross-spectrum $f_{ij\given\V\backslash\lbrace i,j\rbrace}(\mathbf{w}_{st})$ can be regarded as the cross-spectrum of two residual processes $\epsilon_i$ and $\epsilon_j$ computed from $N_i$ and $N_j$. 

Analogously to \eqref{eq:STPPCoh}, rescaling of the partial cross-spectral density function yields the partial spectral coherence function $|R_{ij\given\V\backslash\lbrace i,j\rbrace}(\mathbf{w}_{st})|^2$, 
\begin{equation}
\label{eq:partialspectraSpp}
|R_{ij\given\V\backslash\lbrace i,j\rbrace}(\mathbf{w}_{st})|^2=\frac{f_{ij\given\V\backslash\lbrace i,j\rbrace}(\mathbf{w}_{st})^2}{\left[f_{ii\given\V\backslash\lbrace i,j\rbrace}(\mathbf{w}_{st})f_{jj\given\V\backslash\lbrace i,j\rbrace}(\mathbf{w}_{st})\right]},
\end{equation}
which is also bounded between zero and one. However, different from the ordinary spectral coherence function, this function expresses the linear interrelation of two component processes which remains after the linear effect of all remaining component processes has been removed by orthogonal projection. In this sense, the partial spectral coherence can be understood as the partial correlation defined as a function of frequencies $\mathbf{w}_{st}$ such that $N_i$ and $N_j$ are conditionally independent at all spatial and temporal lags given $N_{\V\backslash\lbrace i,j\rbrace}$ ($\condindep{N_i}{N_j}{N_{\V\backslash\lbrace i,j\rbrace}}$) if $|R_{ij\given\V\backslash\lbrace i,j\rbrace}(\mathbf{w})_{st}|^2$ vanishes at all frequencies $\mathbf{w}$ (cf. \cite{Brillinger1981,Rosenberg1989}).

Having only three distinct components $i,j$ and $k$ under study, the calculation can be simplified though
\[
R_{ij\given k}(\mathbf{w}_{st})=\frac{R_{ij}(\mathbf{w}_{st})-R_{ik}(\mathbf{w}_{st})R_{jk}(\mathbf{w}_{st})}{\sqrt{1-R_{ik}(\mathbf{w}_{st})^2}\sqrt{1-R_{jk}(\mathbf{w}_{st})^2}}.
\]

Besides, an alternative spectral characteristic called the absolute rescaled inverse spectral density function $|d_{ij}(\mathbf{w})_{st}|$ can be calculated from the negative of the partial spectral coherency function, that is $|d_{ij}(\mathbf{w}_{st})|=-R_{ij\given\V\backslash{\lbrace i,j\rbrace}}(\mathbf{w}_{st})$ which measures the strength of the linear partial interrelation between $N_i$ and $N_j$ at frequencies $\mathbf{w}$ (cf. \cite{Dahlhaus2000}).

\subsection{Estimation of partial spectral spatio-temporal density functions}

Whilst Section \ref{sec:partialSPP} formalises the concept of partial point process spectral properties, we now briefly review different computational methods for the calculation of the partial cross-spectrum from empirical data. 

Applying well-known results from the theory of the multivariate normal and \citet[Theorem 8.3.1.]{Brillinger1981}, a first approach which requires the inversion of a $(d-2)\times(d-2)$ matrix is to compute the partial cross-spectrum using the formula
\begin{equation}\label{partial.formula}
f_{ij\given\V\backslash\lbrace i,j\rbrace}(\mathbf{w}_{st})=f_{ij}(\mathbf{w}_{st})-f_{i\V\backslash\lbrace i,j\rbrace}(\mathbf{w}_{st})f_{\V\backslash\lbrace i,j\rbrace\V\backslash\lbrace i,j\rbrace}(\mathbf{w}_{st})^{-1}f_{\V\backslash\lbrace i,j\rbrace j}(\mathbf{w}_{st})
\end{equation}
where \[
f_{i\V\backslash\lbrace i,j\rbrace}(\mathbf{w}_{st})=\left[f_{i1}(\mathbf{w}_{st}), \ldots,f_{ii-1}(\mathbf{w}_{st}), f_{ii+1}(\mathbf{w}_{st}),\ldots,f_{ij-1}(\mathbf{w}_{st}), f_{ij+1}(\mathbf{w}_{st}), \ldots\right]
\]
is a $(d-2)\times 1$ matrix. As an alternative approach, a step-wise procedure for \eqref{partial.formula} can be implemented by recursively applying algebraic operations as described by  \cite{Bendat1978}. However, as the required number of stepwise calculations strictly depends on the number of components under study, this recursive calculus is computationally inefficient in high dimensional settings. Finally, a less computationally intensive approach has been introduced in \cite{Dahlhaus2000} and effectively extended to the spatial domain by e.g. \cite{Eckardt2016} where, under regularity assumptions, the partial spectra can be obtained from the inverse $\boldsymbol{\flat}(\mathbf{w}_{st})$ of the spectral matrix  $\mathbf{f}(\mathbf{w}_{st})$ such that 
\begin{equation}
\label{InverseSpectraRxyz}
R_{ij\given\V\backslash\lbrace i,j\rbrace}(\mathbf{w}_{st})=-\frac{\flat_{ij}(\mathbf{w}_{st})}{\left[ \flat_{ii}(\mathbf{w}_{st})\flat_{jj}(\mathbf{w}_{st})\right]^{\frac{1}{2}}}
\end{equation}
whence 
\begin{equation}
\vert d_{ij}(\mathbf{w}_{st})\vert=\frac{\vert \flat_{ij}(\mathbf{w}_{st}) \vert}{\left[\flat_{ii}(\mathbf{w}_{st})\flat_{jj}(\mathbf{w}_{st})\right]^{\frac{1}{2}}}.
\end{equation} 
We note that expressions for the multivariate-marked case can analogously be computed from the  inverse of the marked cross-spectra $\flat^m_{ij}(\mathbf{w}_{st})$. Likewise, for subsets $X_i$, $\mathbf{X}_J$ and $\mathbf{X}_K$ where $\mathbf{X}_J\cup \mathbf{X}_K\subset\mathbf{X}_{\V\backslash\lbrace i\rbrace}$, replacing  $\flat{ij}(\mathbf{w}_{st})$ by   $\flat_{iK}(\mathbf{w}_{st})$, the inverse of $f_{iK}(\mathbf{w}_{st})$, yields the partial dot-type spectra $f_{iK|K}(\mathbf{w}_{st})$.  Different from ordinary partial spectral characteristics, $f_{iK|J}(\mathbf{w}_{st})$ describes the partial interrelation between component $X_i$ and subset $\mathbf{X}_K$ conditional on subset $\mathbf{X}_J$ which is identical to $f_{ik\given\V\backslash\lbrace i,j\rbrace}$ if and only if $\mathbf{X}_K$ reduces exactly to component $X_k$ and $\mathbf{X}_J=\mathbf{X}_{\V\backslash\lbrace i,k\rbrace}$.

\subsection{Spatial dependence graph model for spatial point processes}\label{Sec:sdgmstpp}

Adopting the ideas of \cite{Eckardt2016} and \cite{Eckardt2019}, we now define the spatio-temporal dependence graph model (henceforth STDGM) aiming to relate the structure of an undirected graph to the partial interrelation structure of a multivariate and multivariate-marked spatio-temporal point pattern. To this end, we identify the vertices of an undirected graph with the components of any such spatio-temporal point process such that an edge between the two vertices $v_i$ and $v_j$ is missing if and only if the component processes $N_i$ and $N_j$ are conditionally uncorrelated after removal of the linear effect of $N_{\V\setminus\lbrace i,j\rbrace}$, e.g. if both components are homogeneous spatio-temporal Poisson processes conditional on all remaining components. This assumption is equivalent to observing a vanishing partial cross-spectrum $f_{ij\given\V\backslash\lbrace i,j\rbrace}(\mathbf{w}_{st})$,  inverse $\flat_{ij}(\mathbf{w}_{st})$, partial spectral coherence function $R_{ij\given \V\backslash\lbrace i,j\rbrace}(\mathbf{w}_{st})$ or absolute rescaled inverse spectral density function $d_{ij}(\mathbf{w}_{st})$ at all frequencies  $\mathbf{w}_{st}$ in space and time. This leads to the following definition of a STDGM.

Let $\mathbf{N}_\V$ be a multivariate or multivariate-marked spatio-temporal point process on $W\times T\subset \mathbb{R}^2\times\mathbb{R}^+$. A spatio-temporal dependence graph model is an undirected graphical model $\G=(\V,\E)$ in which any $v_i\in\V$ encodes a component of $\mathbf{N}_\V$ and $\E=\lbrace \lbrace v_i,v_j\rbrace: R_{ij\given \V\backslash\lbrace i,j\rbrace}(\mathbf{w}_{st})\neq 0\rbrace$ such that
\[
\condindep{N_i}{N_j}{N_{\V\backslash\lbrace i,j\rbrace}}\Leftarrow \lbrace v_i,v_j\rbrace \notin\E.
\]
Hence, a spatial dependence graph model is an undirected graph in which conditional interrelations can be identified from non-missing edges. Precisely, two components $N_i$ and $N_j$ are said to be conditionally uncorrelated at all spatial and all temporal lags after removing the linear effect of all remaining components if the unordered pair $\lbrace v_i, v_j \rbrace , ~i\neq j$ is not in $\E$.

\subsection{Partial characteristics in the spatio-temporal domain}

Adopting Bochner's theorem, different partial spatio-temporal domain characteristics can be computed directly from the partial spectral characteristics through the inverse Fourier transformation.

Whence, for the qualitative marks, applying the inverse transformation yields the spatio-temporal partial complete auto-covariance function, 
\begin{equation}\label{inversekappa1}
\kappa_{ii|V\setminus\lbrace i,j\rbrace}(\mathbf{c},t) 
=\int_{\mathbb{R}^2}\int_{\mathbb{R}}\exp(\imath(\mathbf{w}_\s\T\mathbf{c}+w_\timefreq h))f_{ii|V\setminus\lbrace i,j\rbrace}(\mathbf{w}_\s,w_\timefreq)dw_\timefreq d\mathbf{w}_\s 
\end{equation}
and the partial complete cross-covariance function
\begin{equation}\label{inversekappa2}
\kappa_{ij|V\setminus\lbrace i,j\rbrace}(\mathbf{c},t) =\int_{\mathbb{R}^2}\int_{\mathbb{R}}\exp(\imath(\mathbf{w}_\s\T\mathbf{c}+w_\timefreq h))f_{ij|V\setminus\lbrace i,j\rbrace}(\mathbf{w}_\s,w_\timefreq)dw_\timefreq d\mathbf{w}_\s.
\end{equation}

Recapitulating that $\kappa_{ij}=\zeta_{ij}$, notice that \eqref{inversekappa2} is the partial cross-covariance density function $\zeta_{ij|V\setminus\lbrace i,j\rbrace}$. Further, the partial correlation $\cor_{ij|V\setminus\lbrace i,j\rbrace}$ and the partial scaled covariance density function $\xi_{ij|V\setminus\lbrace i,j\rbrace}$ are then defined as

\[
\cor_{ij|V\setminus\lbrace i,j\rbrace}((\mathbf{s},t),(\mathbf{s}',t'))= (d\mathbf{s}\times t,d\mathbf{s}'\times t')^\frac{1}{2}\frac{\zeta_{ij|V}((\mathbf{s},t),(\mathbf{s}',t'))}{(\lambda_i(\mathbf{s},t)\lambda_j(\mathbf{s}',t'))^\frac{1}{2}}
\]

and
\[
\xi_{ij|V\setminus\lbrace i,j\rbrace}((\mathbf{s},t),(\mathbf{s}',t'))=\frac{\zeta_{ij|V}((\mathbf{s},t),(\mathbf{s}',t'))}{(\lambda_i(\mathbf{s},t)\lambda_j(\mathbf{s}',t'))^\frac{1}{2}}.
\]

Likewise, similar partial point process characteristics can be computed from the inverse transformation of multivariate-marked as well as dot-type partial cross-spectra expression yielding a rich toolbox of novel numerical summary characteristic for spatial-temporal point  data. E.g., for the multivariate-marked case, inverse transformation of the marked auto- and cross-spectral density functions yield the partial auto-type mean product of marks $U_{ii|V\setminus\lbrace i,j\rbrace}(\mathbf{c},t)$ and the partial cross-type mean product of marks $U_{ij|V\setminus\lbrace i,j\rbrace}(\mathbf{c},t)$, respectively, which properties and interpretation has not been investigated yet and highly welcomes deeper evaluations in future research. 
 
\section{Multivariate spatio-temporal crime data}

This section covers the application of the STDGM to a spatio-temporal crime dataset provided under the Open Government Licence by the British Home Office for London and  has been downloaded from {\url{http://data.police.uk/data/}}. This data contains the longitude and latitude for a set of $14$ pre-classified crime categories at street-level, either within a one mile radius of a single point or within a custom area of a street recorded by the Metropolitan Police as well as the month of occurrence for each single crime event. For our analysis we preselected all crime events which have been collected within a four-month period from April to July 2016 yielding a sample of $343427$ single crime events from which $77139$ events have been recorded in April, $86915$ in May, $85761$ in June and, lastly, $93612$ in July. Finally, excluding any duplicated events our data reduces to $127328$ events in total.

To give a first impression about the temporal variation of the spatio-temporal point pattern, the numbers of crimes and different numerical summary characteristics per month have been computed. The monthly numbers and spatial first-order intensity functions per crime category are reported in Table \ref{tab:CrimeStpp}. Inspecting
Table \ref{tab:CrimeStpp}, we found that most of the crime events have been categorised as anti-social behaviour and also violence and sexual offences,  while possession of weapons appeared least often. Further, an increase in numbers of cases of anti-social behaviour from April to July can be observed contrasting with a decrease in numbers of cases of violence and sexual behaviour in the same period. The intensity reports the mean number of events per unit area of the region considered in London.

\begin{table}
\centering
\scalebox{0.8}{
\begin{tabular}{lrrrr}
  \hline\hline
\multicolumn{1}{l}{crime}& \multicolumn{1}{c}{April}  &\multicolumn{1}{c}{May} &\multicolumn{1}{c}{June}& \multicolumn{1}{c}{July} \\
  \hline	
Anti-social behaviour	&$9824~(494.079)$&$11420~(529.492)$&$11570~(388.770)$&$ 13280~(675.639)$\\
Bicycle theft	&$800~(40.234)$&$1008~(46.736)$&$968~(32.526)$&$ 1081~(54.997)$\\
Burglary	&$3440~(173.008)$&$3257~(151.012)$&$3204~(107.660)$&$ 3040~(154.664)$\\
Criminal damage and arson	&$2738~(137.702400)$&$2940~(136.314000)$&$2579~(86.659)$&$ 2694~(137.061)$\\
Drugs	&$1188~(59.748)$&$1108~(51.373)$&$1056~(35.483)$&$ 1056~(53.725)$\\
Other crime	&$159~(7.997)$&$150~(6.955)$&$143~(4.805)$&$ 147~(7.479)$\\
Other theft	&$2673~(134.433)$&$2816~(130.565)$&$2799~(94.051)$&$ 2682~(136.451)$\\
Possession of weapons	&$98~(4.929)$&$99~(4.590)$&$137~(4.603)$&$ 106~(5.393)$\\
Public order	&$1049~(52.757)$&$1079~(50.028)$&$1164~(39.112)$&$ 1205~(61.306)$\\
Robbery	&$444~(22.330)$&$510~(23.646)$&$557~(18.716)$&$ 502~(25.540)$\\
Shoplifting	&$380~(19.111)$&$340~(15.764)$&$378~(12.701)$&$ 324~(16.484)$\\
Theft from the person	&$422~(21.224)$&$424~(19.659)$&$549~(18.447)$&$ 459~(23.352)$\\
Vehicle crime	&$2983~(150.024)$&$2994~(138.818)$&$2907~(97.680)$&$ 2746~(139.707)$\\
Violence and sexual offences	&$4024~(202.379)$&$3946~(182.958)$&$3905~(131.214)$&$ 3777~(192.160)$\\
\hline
\hline
\end{tabular}}
\caption[Numbers and average intensity of spatio-temporal crime pattern]{Number of events and average spatial intensity per crime category per month: Monthly numbers of $14$ different crime categories recorded from April 2016 to July 2016 in London and first-order spatial intensity functions per crime type in brackets. \label{tab:CrimeStpp}}
\end{table}

\begin{figure}[h!]
\begin{center}
\includegraphics[width=0.90\linewidth]{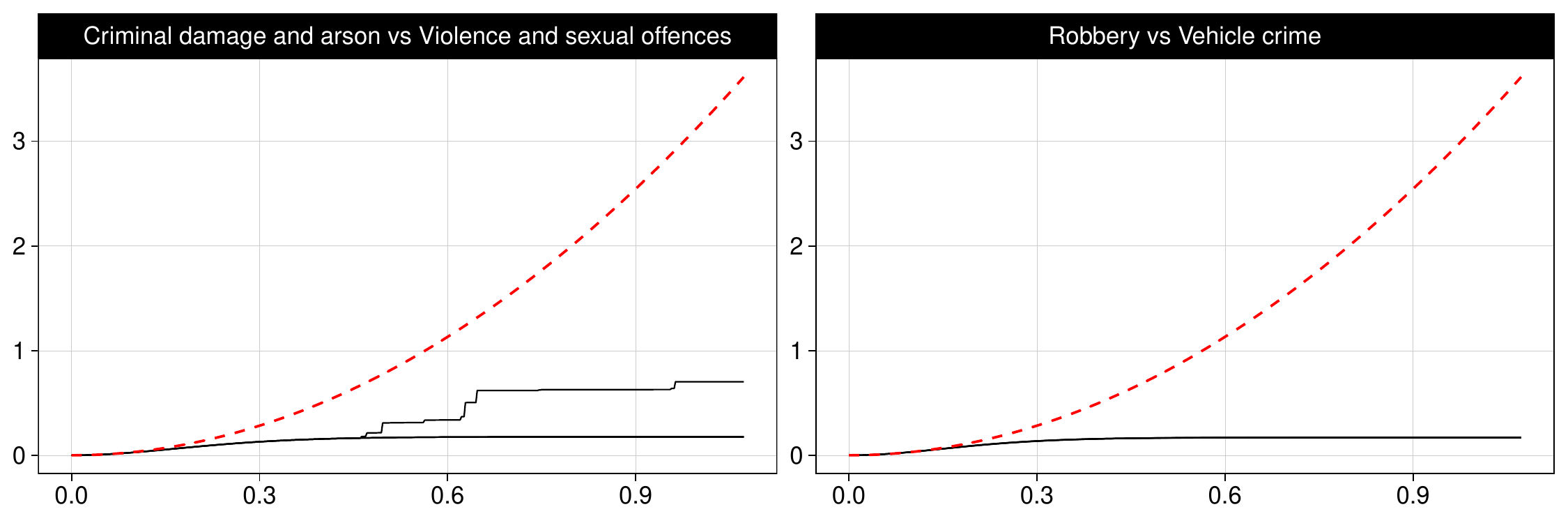}
\caption{Bivariate spatial $K$-function per month for the pairs Criminal damage and arson vs violence (left) and Robbery vs vehicle crime (right). Each black line corresponds to each month, and the dashed red line represents the theoretical value under independence amongst types.}
\label{Fig:kfct}
\end{center}
\end{figure}

We note that this current dataset contains only four temporal instants, and this lack of temporal information basically prevents from running a formal spatio-temporal analysis using second-order characteristics as in Section 2. In addition, so many points in space for each type of crime make computational burden when using for example the spatial $K$-function. Two more points are in order. The crime data happens on the streets of London, and the network structure is key in the spatial structure of the events. Note that there is a large hole within the spatial region having to do with the network itself. This is not considered into account in the functions shown in Section 2 for a more classical spatio-temporal analysis. A second aspect is that this classical approach does not consider conditional relationships, and can only measure global bivariate cross-relationships when marks are discrete. Due to these number of drawbacks, we considered the bivariate spatial $K$-function per month for the pairs (Criminal damage and arson vs Violence and sexual offences) and (Robbery vs Vehicle crime). The corresponding $K$-functions per month are displayed in Figure \ref{Fig:kfct}, where the Poisson line is also depicted. 
In addition, they show a sort of regular, inhibitory structure between the two types of crimes for each pair considered. It is clear that time is not rightly considered here, and the analysis of the spatial structure neglects the remaining information from the other types. These facts motivate our new graphical modelling approach as follows.

\subsection{Cross-sectional graphical modelling}
\begin{sidewaysfigure}
  \centering
\subfloat[]{\label{april}
    \makebox{\includegraphics[scale=.39]{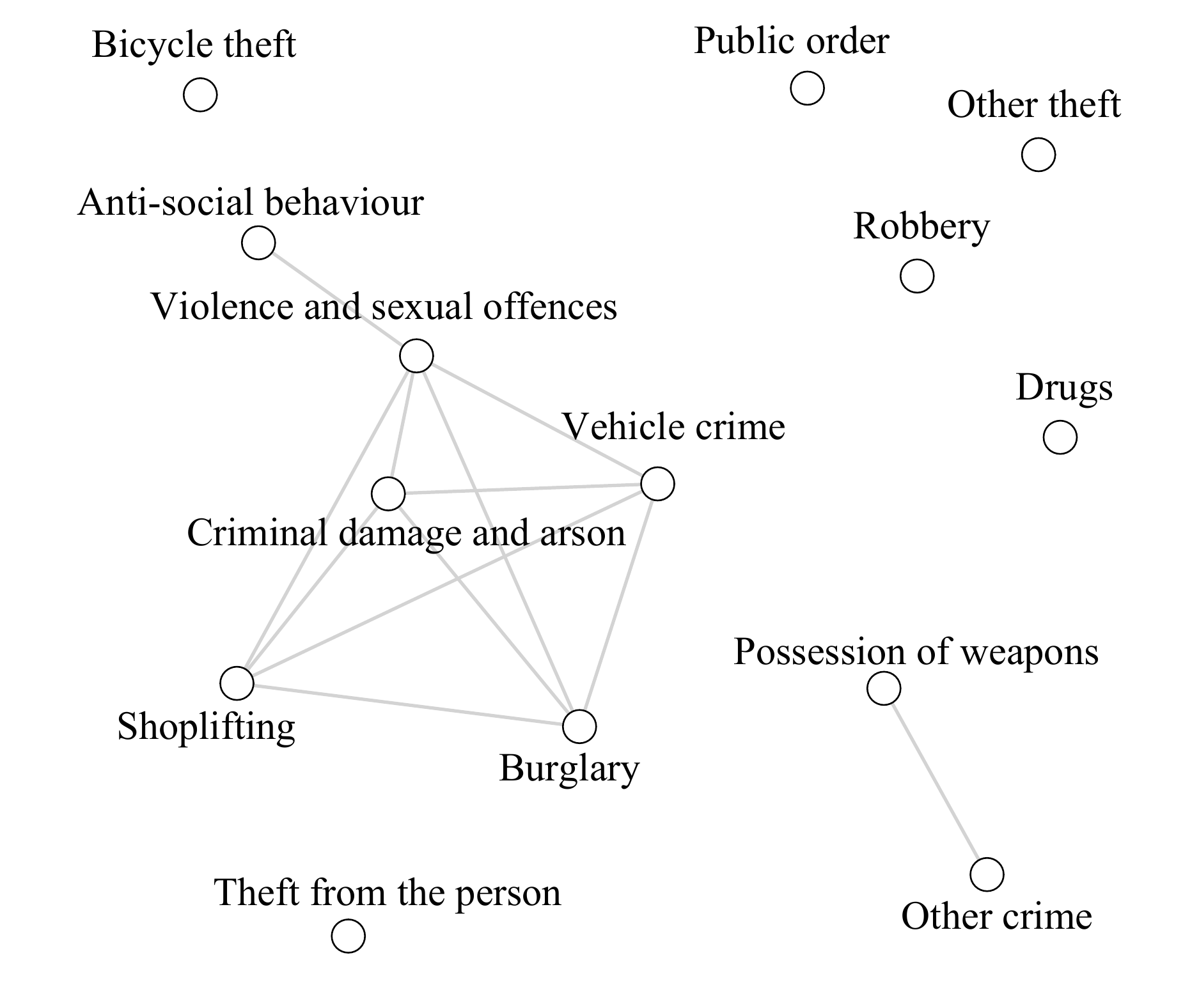}}}
    \hspace{0.8cm}
\subfloat[]{\label{may}
    \makebox{\includegraphics[scale=.39]{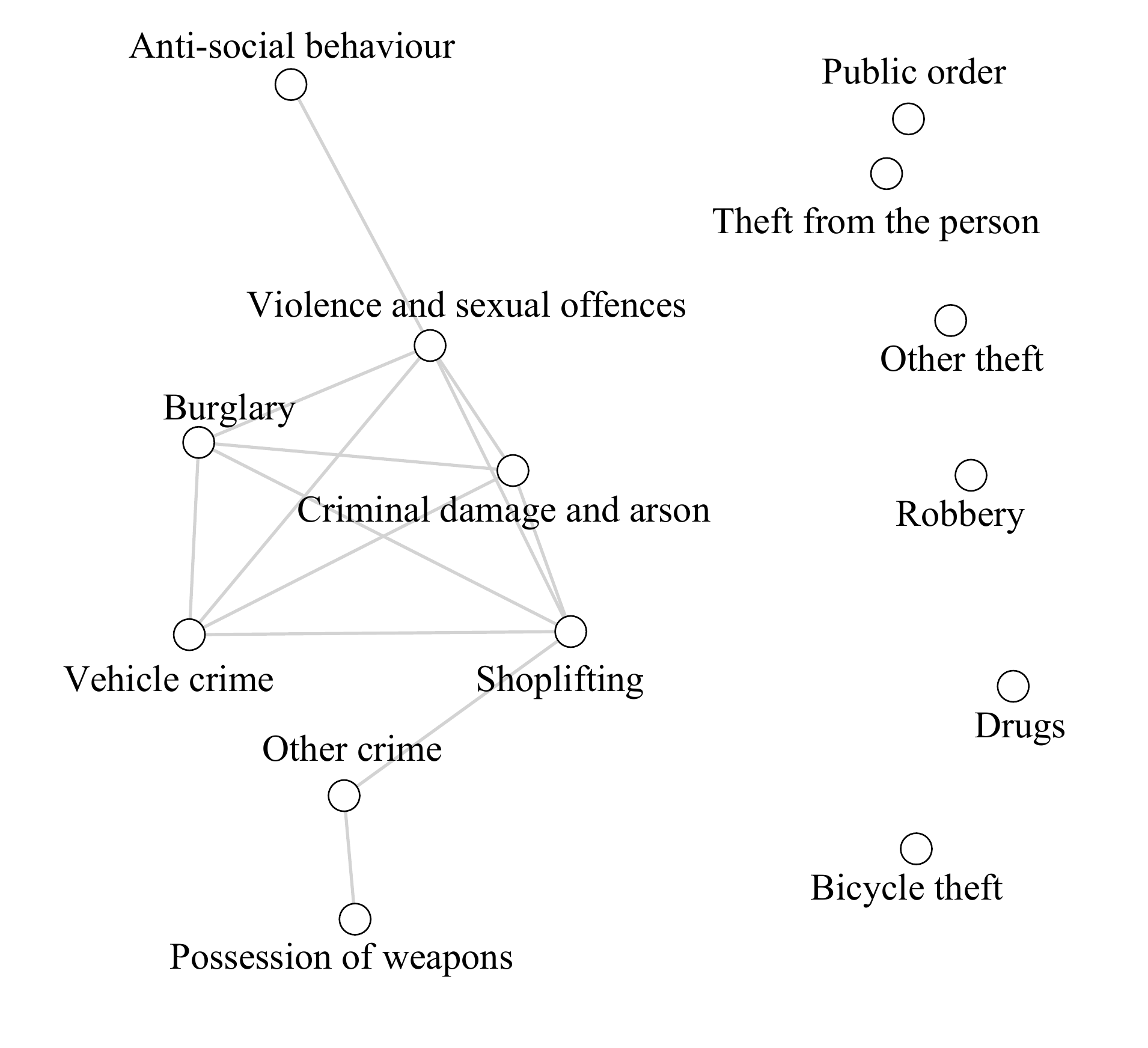}}}\\
\subfloat[]{\label{june}
    \makebox{\includegraphics[scale=.39]{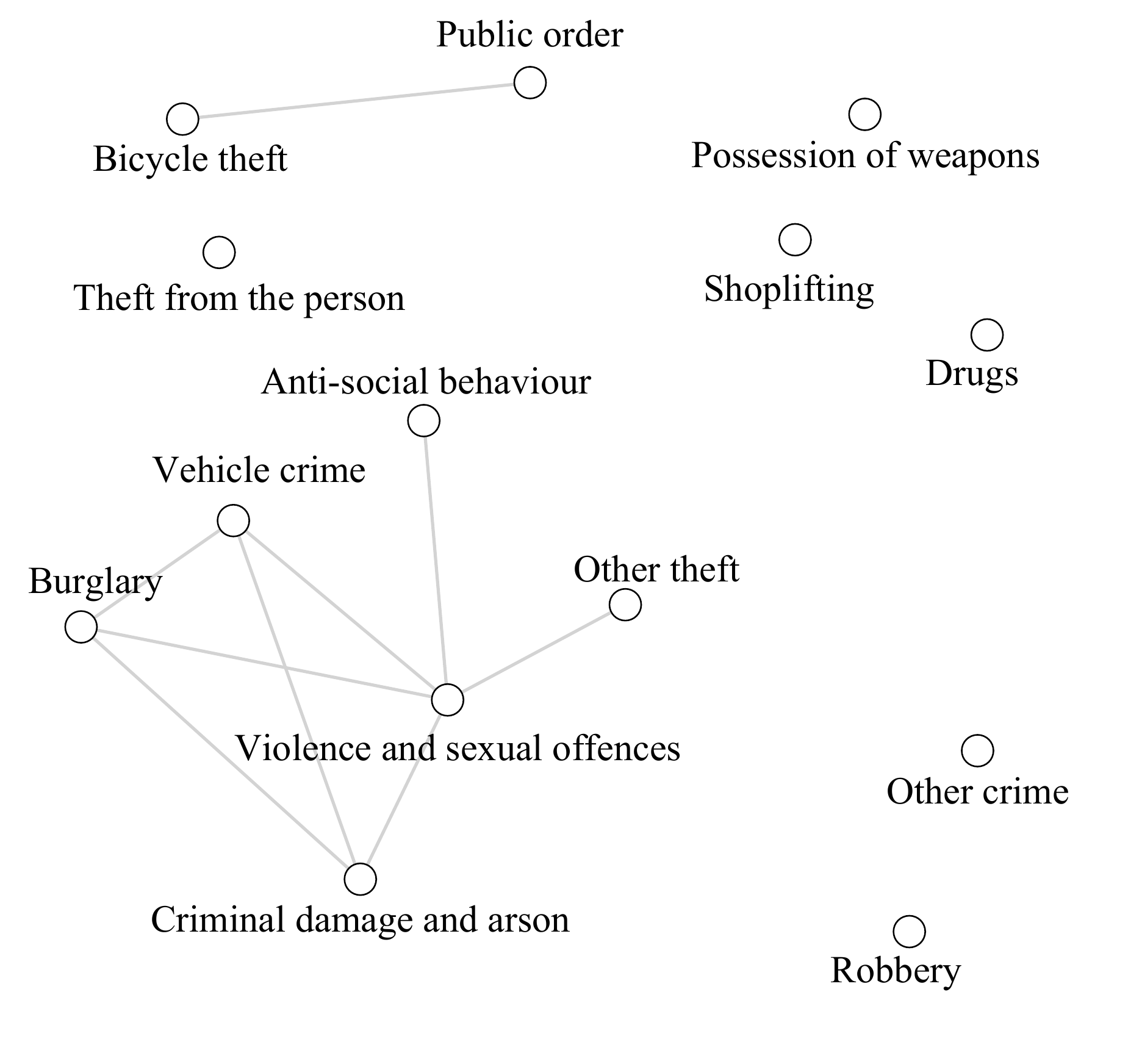}}}
      \hspace{0.8cm}
\subfloat[]{\label{july}
    \makebox{\includegraphics[scale=.39]{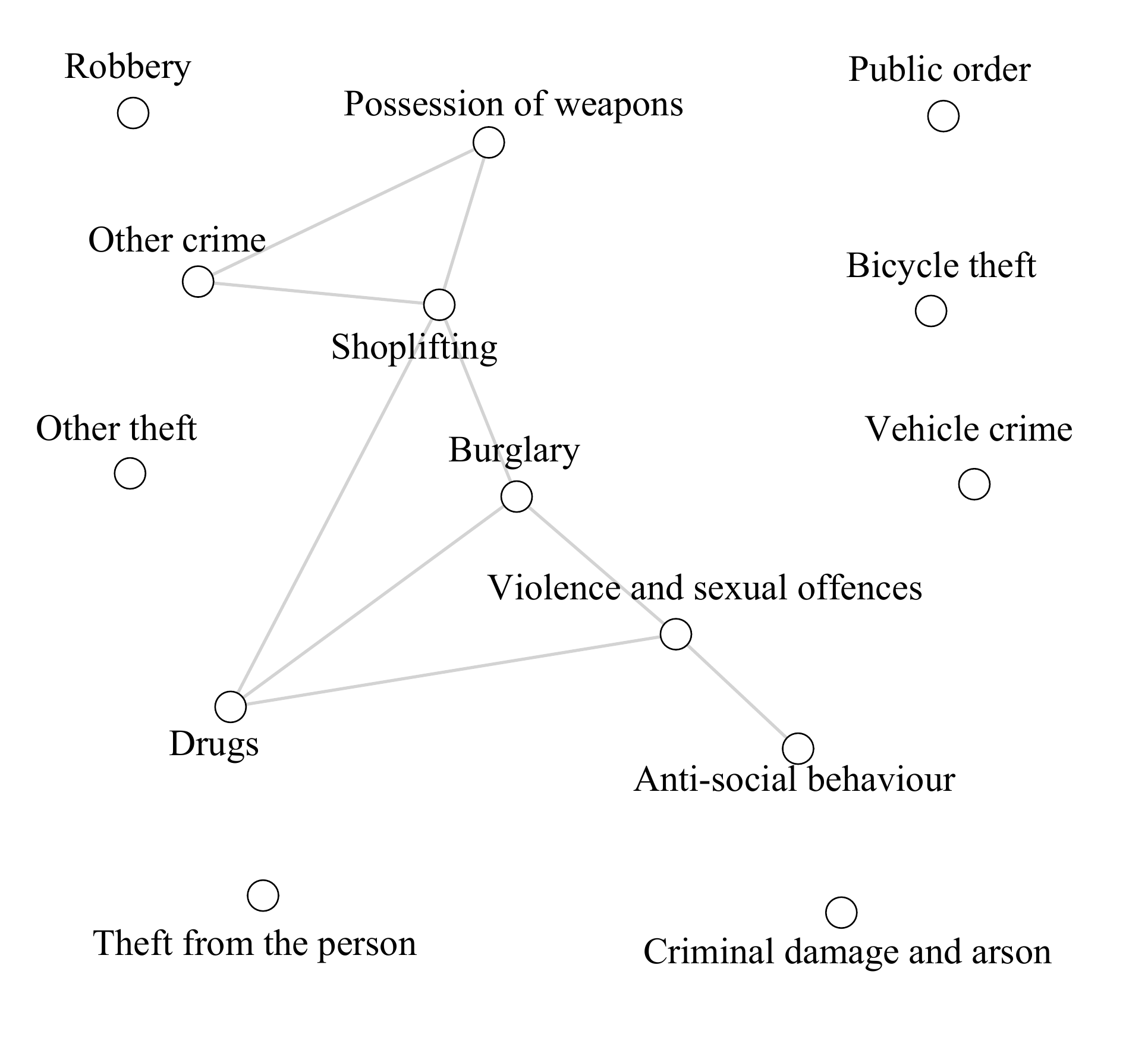}}}
\caption[Spatial dependence graph models per month for $14$ different crime categories recorded in London for a threshold level of $\xi=0.6$]{\label{Fig:CrimeMOnthly} Spatial dependence graph models per month for $14$ different crime categories recorded in London for a threshold level of $\xi=0.6$:  (a) April 2016, (b) May 2016, (c) June 2016 and (d) July 2016.}
\end{sidewaysfigure}

To investigate the structural interrelations among the $14$ crime categories from a cross-sectional perspective and to evaluate the findings of the STDGM, we first discuss the SDGMs computed for each month separately. To this end, we split the data per month into four subsets and computed separate spatial auto- and cross-periodograms. To control for possible variation in strength of the partial interrelations between different pairs of crimes,  we considered a threshold level of $\xi=0.6$ to discover partial interrelations with an intermediate effect size. That is, for each SDGM an edge is drawn between the nodes $i$ and $j$ if the supremum of the empirical  absolute rescaled inverse spectral density function for components $i$ and $j$ equals or exceeds $\xi$ for at least one frequency $\mathbf{w}$ for $p = 0, 1, \ldots, 16$ and $q = - 16, \ldots, 15$ for the particular month. In this case, the point distributions of the components $i$ and $j$ recorded for a particular time $t, t=1,\ldots,4$ are said to be interrelated. The resulting monthly SDGMs are depicted in Figure \ref{Fig:CrimeMOnthly}.

Looking at these plots, we found $6$ isolated nodes in Figures \ref{april} to \ref{june} and $7$ isolated nodes in Figure \ref{july}. For Figure \ref{april} and Figure \ref{may}, we observed an identical set of isolated nodes (public order, other theft, robbery, drugs, theft from the person, bicycle theft) while different sets of isolated nodes are shown in Figure \ref{june} (possession of weapons, shoplifting, drugs, other crime, robbery and theft from the person) and  Figure \ref{july} (public order, bicycle theft, vehicle crime, criminal damage and arson, theft from the person, other theft and robbery). For all isolated nodes, we concluded that none of these crime patterns are interrelated to any alternative crime pattern included from a cross-sectional perspective. We note that comparing the isolated nodes over the complete period under study, only two crimes remain isolated throughout the four months, namely robbery and theft from the person, while most alternative isolated nodes appeared in, at most, three SDGMs. Further, other crime and shoplifting (resp. criminal damage and arson and vehicle crime) are not interconnected to any alternative crime  in June (resp. July).  We outline that the isolated nodes would imply that the distributions of point locations of any of these crimes obey complete spatial randomness conditional on all alternative crime patterns included by definition in the spatial dependence graph model.

Besides these isolated nodes, several subgraphs can be identified. Inspecting the upper panel, two subgraphs are shown in Figure \ref{april} (a $2$-node and a $6$-node subgraph) while all alternative crimes are joined in only one $8$-node subgraph in Figure \ref{may}. Turning to the lower panel, we again observed a $2$-node and a $6$-node subgraph for June (Figure \ref{june}) whereas all non-isolated nodes form a subgraph in Figure \ref{july}. Further, looking at all four spatial dependence graph models, we found that anti-social behaviour is only directly connected with violence and sexual offences which implies that anti-social behaviour is conditionally independent of all remaining crimes given violence and sexual offences from a purely spatial perspective for all months. Comparing this finding with Table \ref{tab:CrimeStpp}, we found that anti-social behaviour, which appeared most often in all four months,  is directly connected to the second most often crime category.
At the same time, we also found that burglary is linked to violence and sexual offences throughout the complete period from a cross-sectional perspective. Except for June (Figure \ref{june}), a direct interrelation can also be detected for possession of weapons and other crime. Interestingly, unlike anti-social behaviour and violence and sexual offences, we observed an opposite relation between the numbers of possession of weapons and other crime throughout all four months with high numbers for other crime while possession of weapons appeared least often.

\subsection{Spatio-temporal dependence graph model results}

We now discuss the results of the STDGM computed from the crime data over the four-month period. Unlike the cross-sectional analysis, the estimation of the STDGM is related to Fourier transformations of the spatial frequencies over time. We note that, as time is assumed to be recorded in equidistant steps in discrete time, the interval length has an important impact on the estimation of the STDGM. As pointed out by  \cite{Didelez2003}, large  intervals result in a marginalisation over time and information on short-term dependencies between different components might be lost. At the same time, additional correlation could emerge due to common causes which occurred in the meantime.

\begin{figure}[h!]
\begin{center}
\includegraphics[scale=0.5]{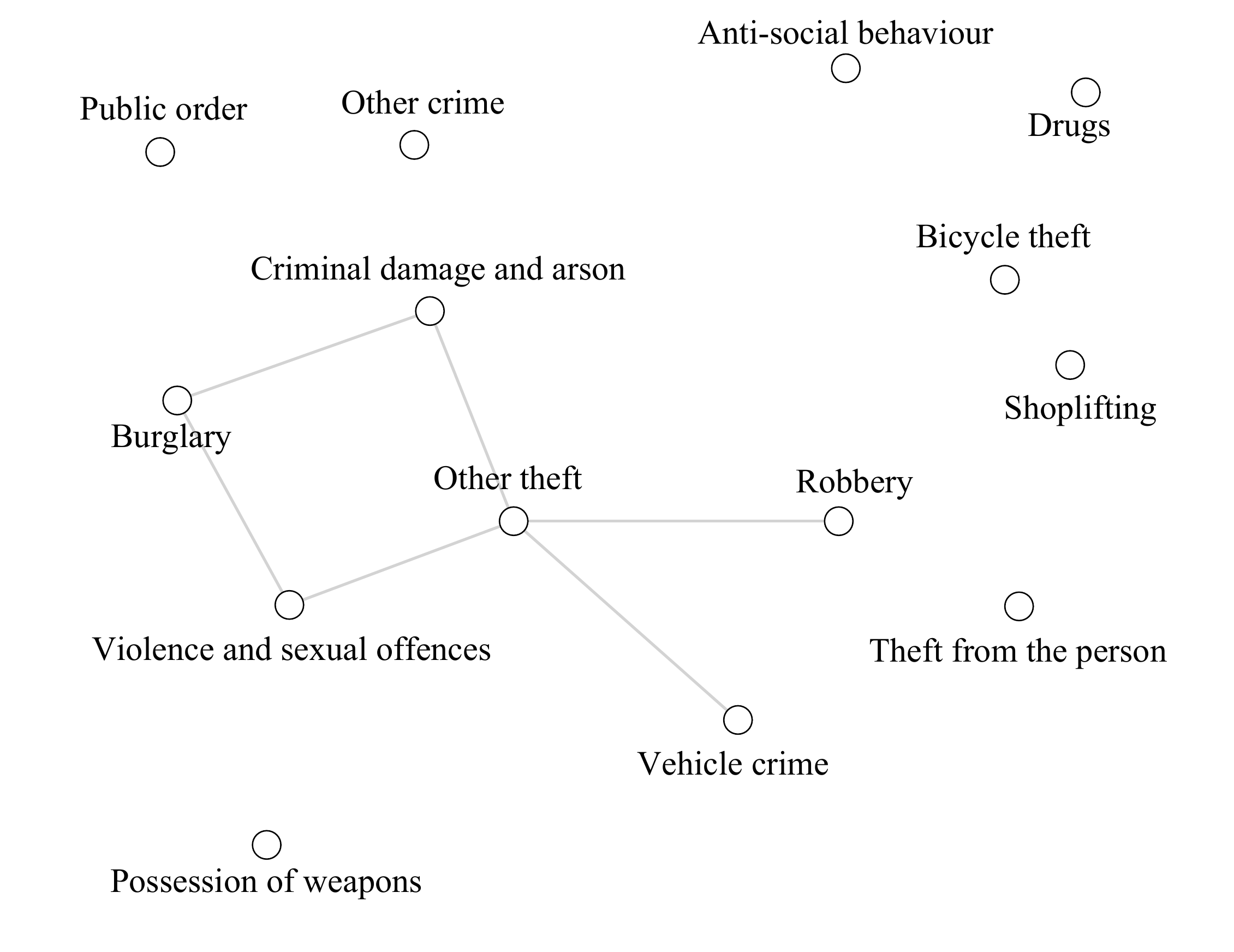}
\caption[Spatio-temporal dependence graph model for $14$ types of crime recorded in London over a four-month period for a threshold level of $\xi=0.6$.]{Spatio-temporal dependence graph model for crime data recorded in London for a four-month period from April 2016 to July 2016 for a threshold level of $\xi=0.6$.}
\label{Fig:CrimeTS}
\end{center}
\end{figure}

To control for possible variation in strength of the partial interrelations between different pairs of crimes over time, we consider a threshold level of $\xi=0.6$ in order to detect conditional partial interrelations with an intermediate effect size such that an edge is drawn between the nodes $i$ and $j$ if the supremum of the empirical  absolute rescaled inverse spectral density function for components $i$ and $j$ equals or exceeds $\xi$ for at least one frequency $\mathbf{w}_{st}$ for $p=0,\ldots, 16,~q=-15,\ldots,16$ and $\timefreq=-2,\ldots,2$. That is, edges indicate that the strength of the linear partial interrelation between two component processes is greater than or equal to $\xi=0.6$ over all four months.  In this particular case, the spatial point distributions of the components $i$ and $j$ are said to be interrelated over time. To state this in a different manner, as the STDGM is defined through the Fourier transform of the spatial frequencies $(p,q)$ at times $t$, edges represent periodicities of $(p,q)$ over time. The resulting STDGM is depicted in Figure \ref{Fig:CrimeTS}.

Inspecting the STDGM, eight isolated nodes (public order, other crime, anti-social behaviour, drugs, bicycle theft, shoplifting, theft from the person, possession of weapons) and one 6-node subgraph can be identified. For the isolated nodes,
we concluded that none of these crimes are interrelated to any alternative crime included over the four-month period. Further, comparing the isolated nodes of the STDGM with the four SDGMs depicted in Figure \ref{Fig:CrimeMOnthly}, no link is drawn joining anti-social behaviour and violence and sexual offences for the spatio-temporal case while this interrelation occurred in all cross-sectional plots from April to July 2016. This implies that, although both crimes are interrelated from a cross-sectional perspective, no periodic structures can be found over monthly time intervals. At the same time, as for the analysis of time series, periodicities might be detected for alternative interval lengths.

Turning to the 6-node subgraph, we observed that the spatio-temporal patterns of robbery as well as of vehicle crime are conditionally independent of all remaining crime patterns given the spatio-temporal distribution of other theft. Interestingly, we also observed that burglary is again linked to violence and sexual offences which also holds for the purely spatial dependence structures as depicted in Figures \ref{april} to \ref{july}. This implies that the interrelations of both crimes  are also periodic over monthly time intervals. We emphasise that these findings would  not have been detected by the classical spatial analysis as conditioning nor partialisation are not able in their case. 

\section{Multivariate spatio-temporal point patterns with quantitative marks}

\subsection{Multivariate-marked spatio-temporal forestry data}

This section discusses the STDGM computed from the Duke Forest data. The Duke Forest data contains information on the longitude and latitude, the diameter at breast height (henceforth DBH) as well as elevation and disease characteristics for $71$ different botanic tree species  recorded for different years in the Duke Forest.  The DBH is the diameter of a tree measured at height of $1.30$ metres above the ground level and is a common method used for estimation of the amount of timber and the age of a single tree. The Duke Forest is owned and managed by the Duke University and covers an area of 7000 acres of forested land as well as open fields located in Durham, Orange and Alamance counties in North Carolina (USA). In total, $14992$ distinct trees reported in a wide format data sheet are repeatedly surveyed at a yearly, mostly biennial, basis within the temporal period from 2000 to 2014.  We note that not all trees are investigated at each single time step yielding time-varying sets of  non-missing DBH values over the complete temporal window.
Focussing on the particular years 2000, 2002, 2004, and 2006, we have $38074$ individual tree location with non-missing DBH information collected over the four-years period. From this subset we finally selected a sample of $20293$ individual locations, where $n_1=5338$, $n_2=5052$, $n_3=4769$ and $n_4=5134$ including eight different tree species requiring that for each single species approximately 100 point locations are reported for each time step.  

\begin{figure}[h!t]
\begin{center}
\includegraphics[scale=0.5]{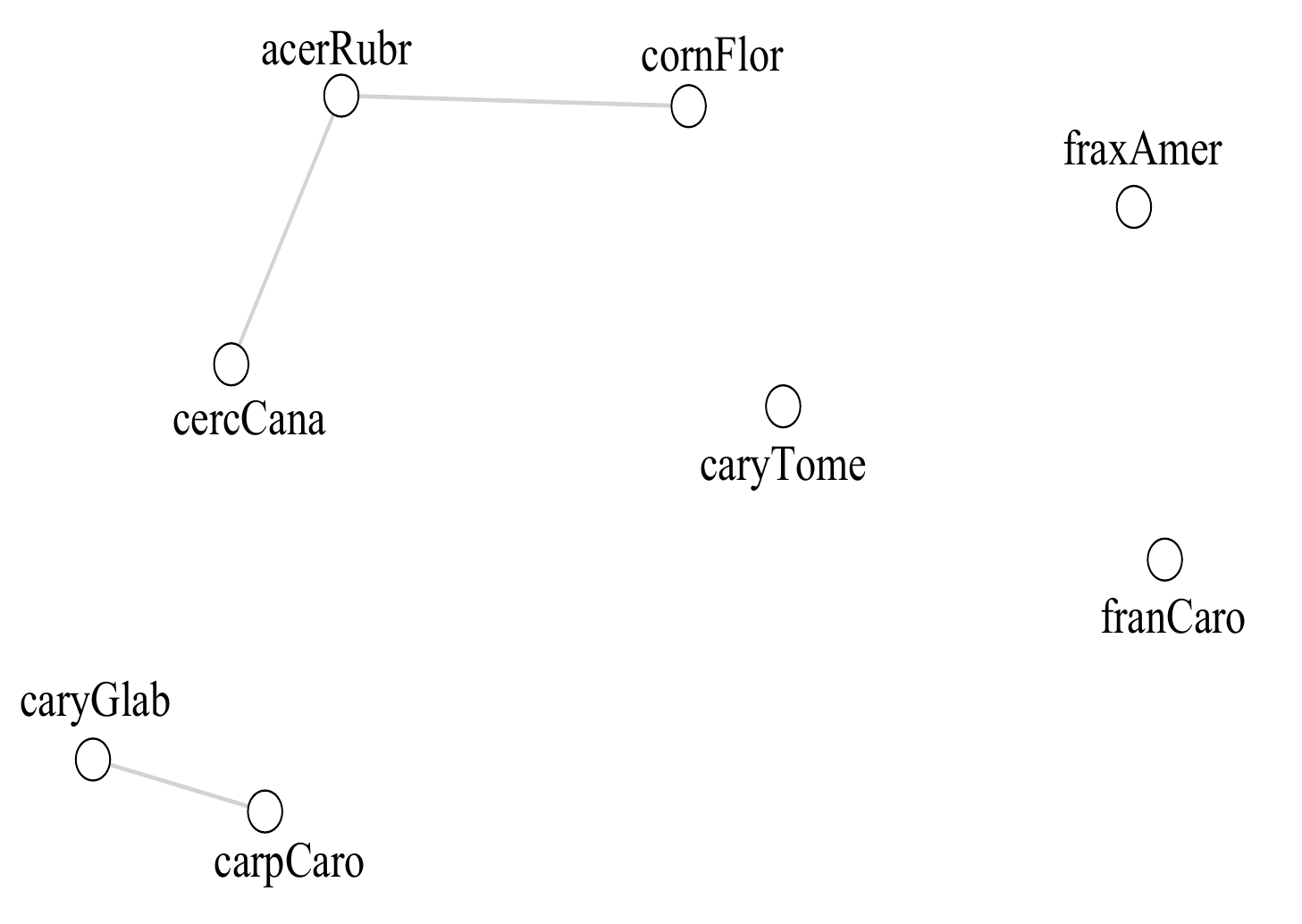}
\caption{Spatio-temporal dependence graph model for a subset of eight tree species recorded in the Duke forest data and biennial DBH values as quantitative marks, for a threshold value of $\xi=0.5$.}
\label{Fig:duke}
\end{center}
\end{figure}

To control for possible variation in strength of the partial interrelations between different pairs of marked locations over time, we consider a threshold level of $\xi=0.5$ in order to detect conditional partial interrelations with an intermediate effect size such that an edge is drawn between the nodes $i$ and $j$ if the supremum of the empirical  absolute rescaled inverse spectral density function for components $i$ and $j$ equals or exceeds $\xi$ for at least one frequency $\mathbf{w}_{st}$ for $p=0,\ldots, 16,~q=-15,\ldots,16$ and $\timefreq=-2,\ldots,2$. In this respect, edges indicate that the strength of the linear partial interrelation between two component processes is greater than or equal to $\xi=0.5$  over all four years. The resulting STDGM is depicted in Figure \ref{Fig:duke}.

Inspecting the STDGM, three isolated nodes (fraxAmer, caryTome, franCaro)  as well as one pair and one triplet of connected edges can be identified. For the isolated nodes,
we concluded that the quantitative marks and the locations of these particular species are both unrelated to the quantitative marks and the locations of any alternative species included over the complete temporal period. We remind that the present formulation of the STDGM only allows for the detection of linear partial interrelations and potential non-linear interrelation are not captured.  Turning to the pair and triplet of interconnected nodes, we found that the marks and locations of (a) caryGlab and carpCaro as well as (b) cercCana, acerRubrand cornFlor are not interrelated to those of any alternative species in the sample except of those species contained in the corresponding 2-node and 3-node subgraphs. At the same time, focussing on (b) and applying basic graph terminology we found that the marked locations of cercCana and cornFlor are conditionally uncorrelated given those of acerRubrand  as acerRubrand serves as a separator in this subgraph.

\begin{figure}[h!tb]
	\begin{center}
		\includegraphics[width=0.9\linewidth]{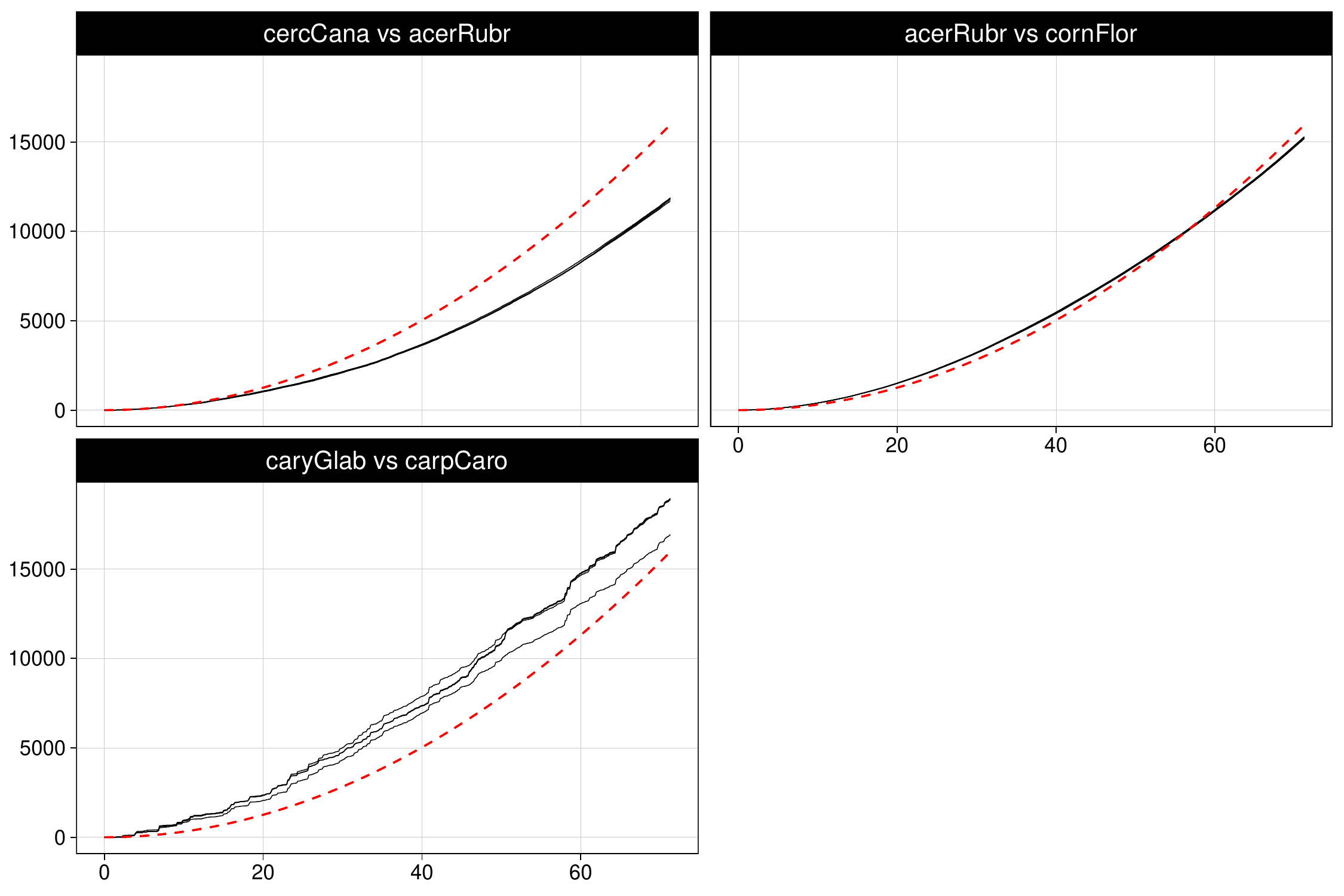}
		\caption{ Bivariate spatial $K$-functions per biennial times for the pairs of species that seem to have dependence through the spatio-temporal dependence graph model. Dashed red lines represent the theoretical value under non-dependence amongst species.}
		\label{Fig:dukeKij}
	\end{center}
\end{figure}

\begin{figure}[htb]
	\begin{center}
		\includegraphics[width=0.99\linewidth]{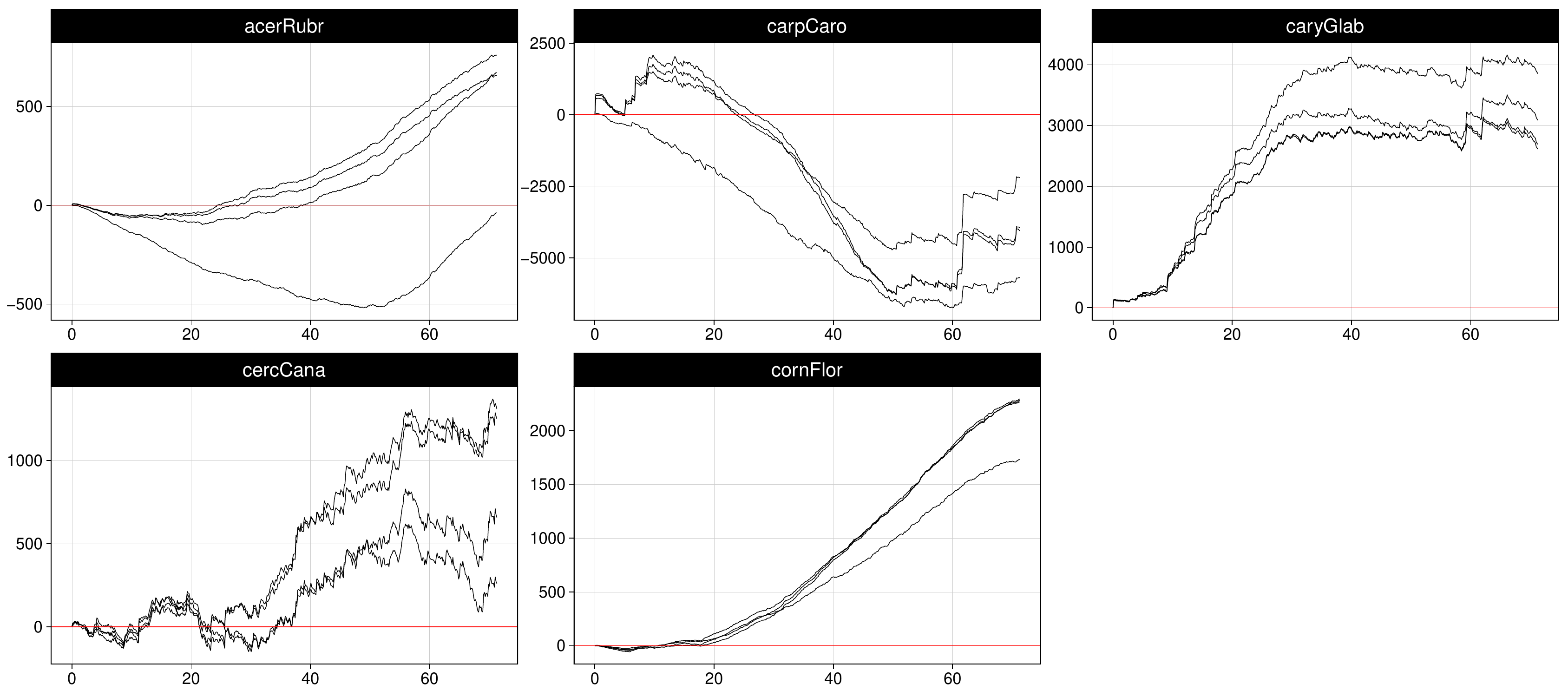}
		\caption{Centred estimates of the mark-weighted $K$-function (by subtracting the $K$-function for the respective unmarked point patterns) of the species of Duke forest that seem to depend on each other. Different lines represent the two-yearly temporal bins and red line represent the independence.}
		\label{Fig:dukeMark}
	\end{center}
\end{figure}

For completeness, we consider the only analysis we can run from the more classical spatio-temporal point of view. Two drawbacks are in order here. One is that as we have only four temporal instants, we can not provide a deep spatio-temporal analysis. Second, we can only show marginal analysis of the whole problem, one with bivariate $K$-functions, and the other with mark weighted $K$-functions, but not an overall analysis. We then report the cross spatial $K$-functions for each temporal bin for all connected types in Figure \ref{Fig:duke} as well as individual mark weighted $K$-functions (see e.g. \citealp{pettinen1992forest}). $K$-functions are displayed in Figures \ref{Fig:dukeKij} and \ref{Fig:dukeMark}.
In all the cases the differences between $\hat{K}_{ij}(r)$ and the benchmark clearly suggest that the components are dependent, which goes in the line found in Figure \ref{Fig:duke}.
The mark-weighted $K$-function for each recorded time is subtracted from its theoretical value under independence between locations and marks, i.e, from the classical $K$-function. We find quite strong indications of a substantial deviation from independence in this case, again reinforcing the results in Figure \ref{Fig:duke}.

\subsection{Multivariate-marked spatio-temporal crime data}

As a second application, the STDGM computed from multivariate-marked crime data is discussed next. This data contains information on a subset of point locations for both property and violent crimes provided as longitude and latitude, the precise date and time of the event, and different attributes of incidents reported in the Analytical Services Application (ASAP) crime report database by the District of Columbia Metropolitan Police Department (MPD). It has been provided under an Open Government Licence and downloaded from \url{https://dcatlas.dcgis.dc.gov/crimecards/}. This data is shared via an automated process where addresses are geocoded to the District's Master Address Repository and assigned to the appropriate street block. To compute a STDGM from this source, we initially extracted the month and year from the original time indication and calculated the duration of police investigation at place in seconds from two additional date and time indications in the data. This duration was then considered as quantitative mark yielding the desired multivariate-marked representation.

\begin{figure}[h!]
\begin{center}
\includegraphics[scale=0.5]{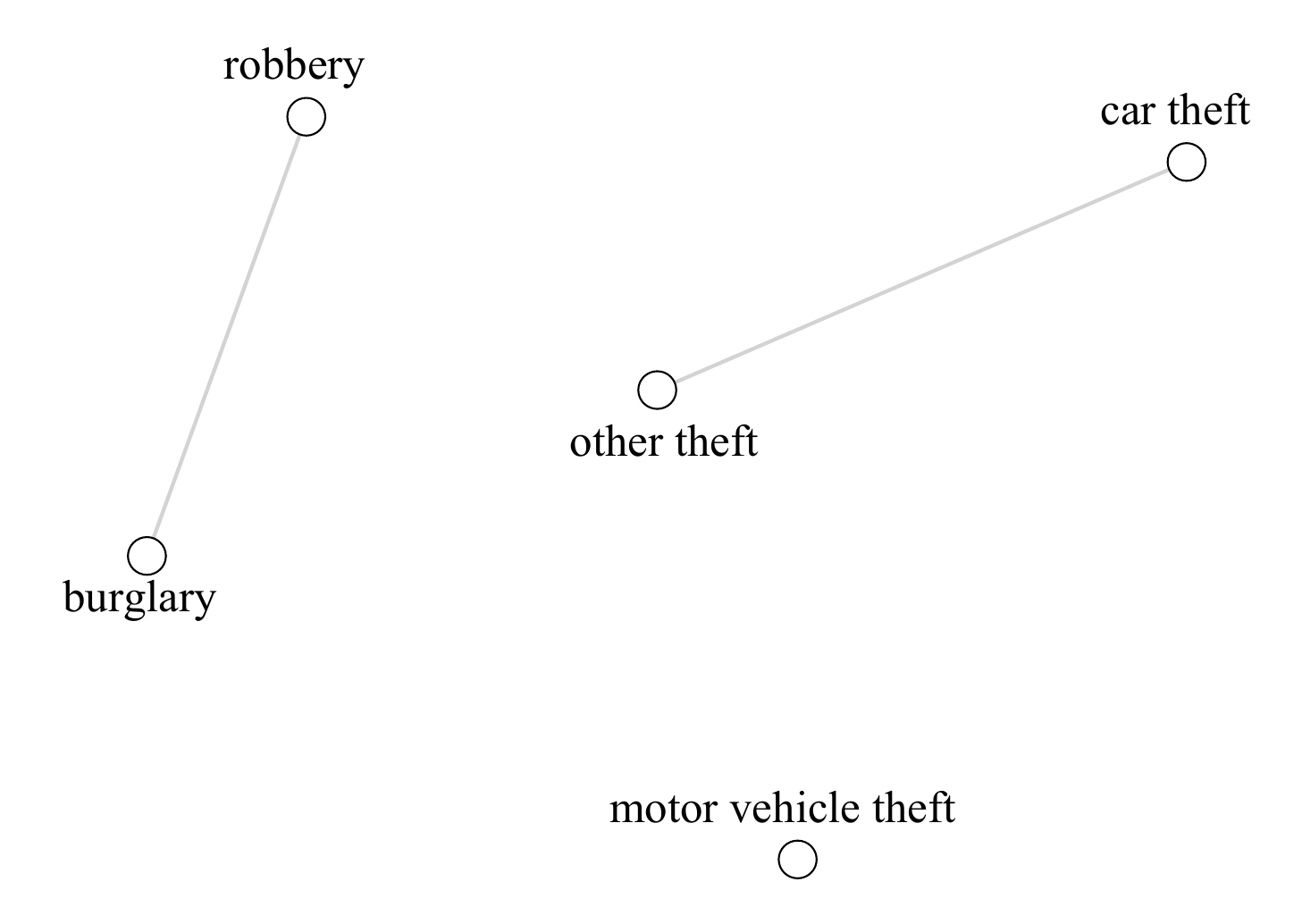}
\caption{Spatio-temporal dependence graph model for different types of crimes recorded in the District of Columbia and investigation period in seconds as quantitative mark for 2019 recorded at monthly  basis for a threshold value of $\xi=0.5$.}
\label{Fig:DCcrimes}
\end{center}
\end{figure}

Restricted to the year 2019 and excluding any cases with either missing date, time or duration information yields a sample of $28999$ events. Taking 12 months into account, we preselected a set of $27680$ point locations from this sample restricted to a set of five distinct pre-specified crime categories. This data then serves as input for the STDGM where we, paralleling the STDGM of the Duke forest data, considered a threshold level of $\xi=0.5$ to control for possible variation in strength of the partial interrelations between different pairs of marked locations over time. This implies that $i$ and $j$ are joint by an edge if the supremum of the empirical  absolute rescaled inverse spectral density function for components $i$ and $j$ equals or exceeds $\xi$ for at least one frequency $\mathbf{w}_{st}$ for $p=0,\ldots, 16,~q=-15,\ldots,16$ and $\timefreq=-2,\ldots,2$. The resulting STDGM is depicted in Figure \ref{Fig:DCcrimes}.

Inspecting the STDGM, we found one isolated node (motor vehicle theft) and two pairs of pairwise interconnected nodes indicating that the marked locations of (a) robbery and burglary, and (b) other theft and car theft meaning theft from car. Interestingly, for (a) the marks and locations  of two different types of crimes (one property and one violent crime) are interrelated to each  other over the complete 12 month period, whereas both crimes of (b) as well as motor vehicle theft are  property related crimes. Neglecting the spatial component and inspecting the number of incident for all five crimes exclusively over the complete period under study, both other theft and car theft show a periodic behaviour with peaks in the summer time while all alternative crimes in the sampled data reflect less fluctuation over time.

\begin{figure}[h!tb]
	\begin{center}
		\includegraphics[width=0.9\linewidth]{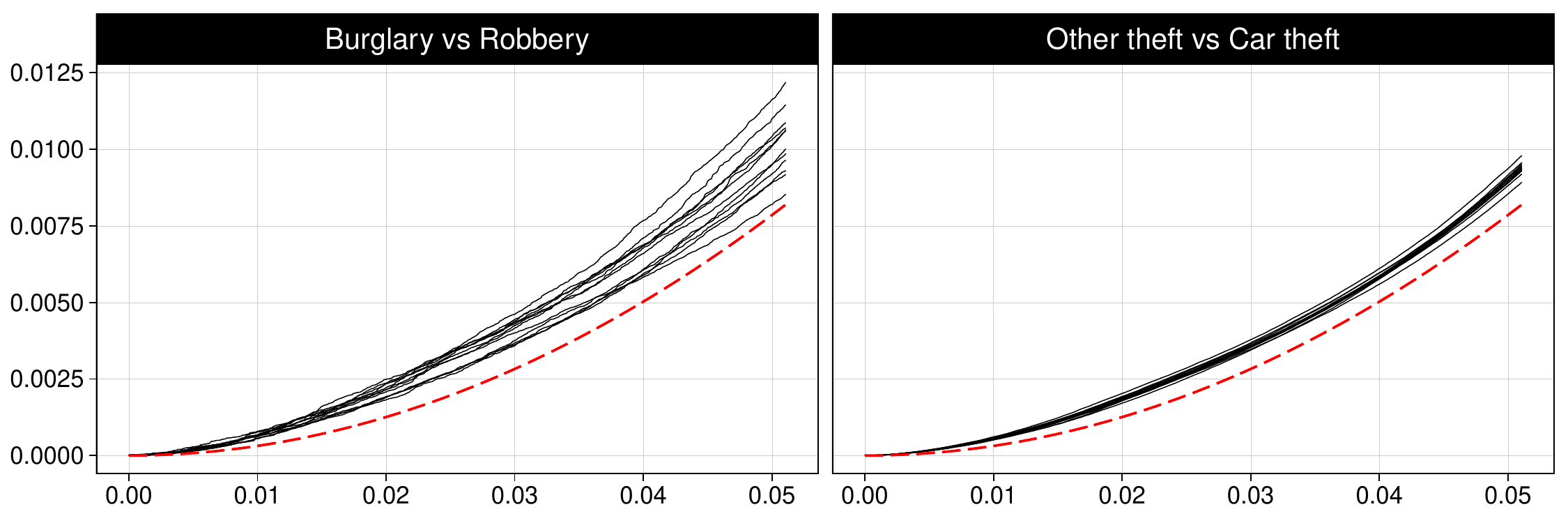}
		\caption{Bivariate spatial $K$-functions per month for the pairs of crimes connected by the spatio-temporal dependence graph model. Dashed red lines represent the theoretical value under non-dependence amongst types of crimes.}
		\label{Fig:crimeKij}
	\end{center}
\end{figure}

Figure \ref{Fig:crimeKij} shows the estimated cross $K$-functions for each month for the connected crimes in Figure \ref{Fig:DCcrimes}.
In the two cases, there are differences between $\hat{K}_{ij}(r)$ and the benchmark, so we can conclude that the components are dependent, reinforcing the results in Figure \ref{Fig:DCcrimes}.
\begin{figure}[htb]
	\begin{center}
		\includegraphics[width=0.9\linewidth]{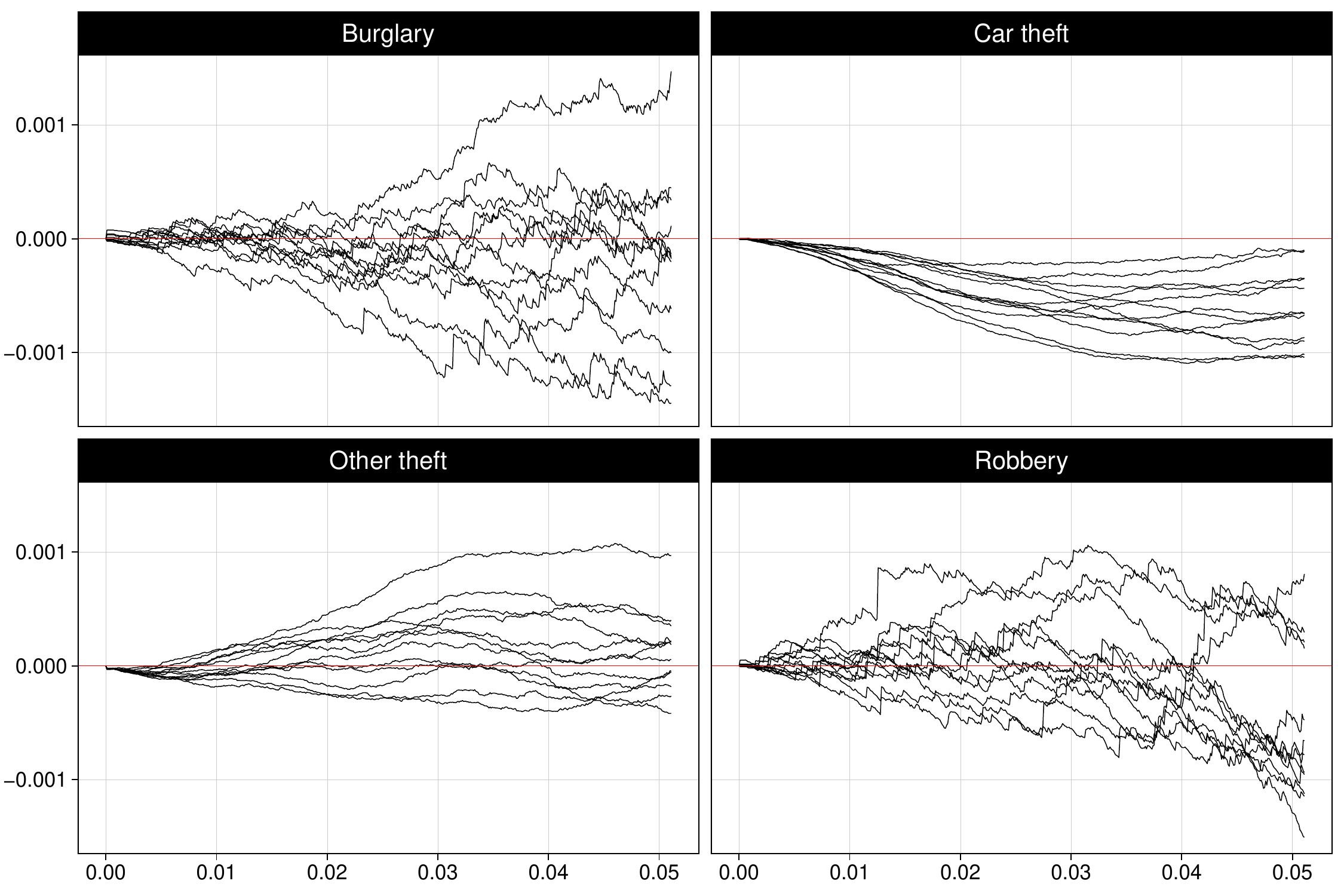}
		\caption{ Centred estimates of the mark-weighted $K$-function of the crime-types in Columbia previously related on each other. Different lines represent the twelve months and red line represent the independence.}
		\label{Fig:crimeMark}
	\end{center}
\end{figure}
Once again, the mark-weighted $K$-function for each recorded time is centred by using its the classical $K$-function. As the lines oscillate around the theoretical value, we can say that there is not enough evidence of substantial deviation from the independence, meaning that the spatio-temporal locations, in this case, neglect somehow the values of the marks. This is different from what we have found in Figure \ref{Fig:DCcrimes}, and this is a misleading result due to the marginal analysis done for the marks only in a spatial context. The STDGM highlights relations that with only a classical method could be missing. 

\section{Conclusions}

The statistical investigation of potential interrelations in marked spatio-temporal point process is an still unresolved and yet highly challenging field of research which has just very recently been started to be explored. When extending classical cross-type characteristics for spatial marked point processes to the spatio-temporal domain, these mark statistics quickly become infeasible and computational burdensome when having large amounts of point locations in time, space or in space-time. To overcome these limitations, the present paper contributes to the multivariate analysis of spatio-temporal point process data by introducing different partial point characteristics and extending the spatial dependence graph model formalism yielding a unified framework for different types of spatio-temporal data including both, purely qualitatively (multivariate) cases and the so-called multivariate-marked spatio-temporal point processes where both qualitative and quantitative information is available for each point location. The proposed graphical model, defined through partial spectral densities characteristics, is highly computationally efficient and reflects in the multivariate-marked case the conditional similarity among sets of spatio-temporal sub-processes of marked points with identical discrete marks.

In addition to the definition of the spatio-temporal dependence graph model, different partial spatio-temporal point  process characteristics are introduced in the frequency spatio-temporal domain which enhance the classical methodology toolbox in multiple ways providing alternative information besides classical univariate and bivariate cross- and dot-type point process statistics as well as  traditional multivariate dimensionality reduction techniques.

Finally, a new class of spectral characteristics is introduced which mirrors the ideas of classical spatial dot-type point point process characteristics to the frequency domain. These dot-type spectral characteristics reflect the interrelation between a particular pattern and one (resp. two) alternative subset (resp. subsets) of components where the pattern of interest is excluded.   

\section*{Acknowledgements}
This research has been partially funded by grants UJI-B2018-04 and MTM2016-78917-R from UJI and the Spanish Ministry of Education and Science.

\bibliographystyle{ecta}
\bibliography{stppgraph}

\end{document}